\documentclass{iucrjournals}
\usepackage{graphicx}
\usepackage{amsmath}
\usepackage{parskip}
\usepackage{float}
\usepackage{fullpage}
\usepackage{siunitx}
\usepackage{xcolor}
\usepackage{hyperref}

\newcommand{\citet}[1]{\protect\citeasnoun{#1}}
\newcommand{\citep}[1]{\protect\cite{#1}}
\usepackage[normalem]{ulem}

\newcommand{\seb}[1]{{\leavevmode\color{blue}{#1}}}
\newcommand{\sebquick}[1]{{\leavevmode\color{blue}{#1}}}
\newcommand{\henry}[1]{{\leavevmode\color{cyan}{#1}}}
\newcommand{\tj}[1]{{\leavevmode\color{green}{#1}}}
\newcommand{\tjquick}[1]{{\leavevmode\color{green}{#1}}}
\newcommand{\tjresolved}[1]{{\leavevmode\color[rgb]{0,0.4,0}{#1}}}

\newcommand{\rr}{\mathbf{r}}
\newcommand{\qq}{\mathbf{q}}
\newcommand{\xtrfacnse}{\chi^{-1}}
\newcommand{\extrapol}{\chi}

\newcommand{\stdxtr}{\sigma_\extrapol}
\newcommand{\occu}{\alpha}
\newcommand{\xtrname}{\chi}

\newcommand{\weight}{w}
\newcommand{\weightfou}{W}
\newcommand{\weightobs}{W_\text{obs}}
\newcommand{\weightxtr}{W_{\xtrname}}
\newcommand{\weightfousf}{\weightfou_\text{SF}}
\newcommand{\weightfouip}{\complex{\weightfou}_\text{IP}}
\newcommand{\weightfougeneral}{\complex{\weightfou}}
\newcommand{\phase}{\varphi}
\newcommand{\complex}[1]{\mathbf{#1}}

\newcommand{\diffnoise}{n_{\Delta}}

\newcommand{\noisediff}{\diffnoise}

\newcommand{\rrneg}{\mathbf{r}_{neg}}

\newcommand{\densdiff}{\Delta \rho}
\newcommand{\densdark}{\rho_0}
\newcommand{\denstarget}{\rho_1}
\newcommand{\densxtr}{\rho_{\xtrname}}
\newcommand{\densmix}{\rho_{mix}}
\newcommand{\denssolvent}{\rho_{solvent}}

\newcommand{\densdiffr}{\densdiff (\rr)}
\newcommand{\densdarkr}{\densdark(\rr)}
\newcommand{\denstargetr}{\denstarget(\rr)}
\newcommand{\densxtrr}{\densxtr(\rr)}

\newcommand{\strucdiff}{\ensuremath{\Delta F}}
\newcommand{\complexstrucdiff}{\ensuremath{\Delta \complex{F}}}
\newcommand{\strucdark}{\ensuremath{F_0}}
\newcommand{\structarget}{\ensuremath{F_1}}
\newcommand{\strucxtr}{\ensuremath{F_{\xtrname}}}
\newcommand{\strucmix}{\ensuremath{F_\mathrm{mix}}}

\newcommand{\subfigref}[2]{%
    \hyperref[#1]{Fig.~(\ref*{#1}#2)}%
}

\newcommand{\FT}{\mathcal F}

\newcommand{\occusim}{\ensuremath{\occu_{sim}}}
\newcommand{\snr}{\mathrm{SNR}}
\newcommand{\gausnoise}{\varepsilon}
\newcommand{\resring}{\mathbf{\mathcal S}}
\newcommand{\rmsddiff}{\ensuremath{\sigma_\Delta}}

\newcommand{\rmsd}{RMSD\@}        
\newcommand{\kweighted}{\textit{k}-weighted}  
\DeclareSIUnit{\us}{\micro\second}    

\newcommand{\combi}{\alpha}
\newcommand{\strucdiffabs}{\Delta|F|}
\newcommand{\deltadark}{{\delta_0}}
\newcommand{\deltadiff}{{\delta_\Delta }}
\newcommand{\densdifffofo}{(\Delta\rho)_\text{obs} }
\newcommand{\densdiffxtr}{(\Delta\rho)_{\xtrname} }
\newcommand{\trueextrapol}{\extrapol_\text{opt}}
\newcommand{\xtrestimate}{\Tilde \extrapol (\rr)}
\newcommand{\densdarkcut}{\densdark^\text{cut}}

\nolinenumbers
\date{}

\begin{document}
\renewcommand{\seb}[1]{}
\renewcommand{\tj}[1]{}
\renewcommand{\sebquick}[1]{}
\renewcommand{\tjquick}[1]{}
\renewcommand{\tjresolved}[1]{}
\renewcommand{\henry}[1]{#1}
\title{Estimating Extrapolation Factors in Time-Resolved Crystallography}
\author[a,b]{
    Sebastian Bielfeldt
}
\author[a,c,d]{
    Henry~N.~Chapman
}
\author[a]{
    Thomas J. Lane
    \IUCrCemaillink{thomas.lane@desy.de}
}

\affil[a]{
    Center for Free-Electron Laser Science CFEL, Deutsches Elektronen-Synchrotron DESY, Notkestr. 85, 22607 Hamburg, Germany
}
\affil[b]{
    Hamburg University of Technology, Institute for Algorithms and Complexity, Hamburg, Germany
}
\affil[c]{The Hamburg Centre for Ultrafast Imaging, Luruper Chaussee 149, 22761 Hamburg, Germany}
\affil[d]{Department of Physics, University of Hamburg, Luruper Chaussee 149, 22761 Hamburg, Germany}

\maketitle
\enlargethispage{\baselineskip}
\begin{abstract}
    Time-resolved crystallography (TRX) captures macromolecules in action, but a triggered crystal contains a mixture of triggered proteins in an unknown state and untriggered proteins in the reference state.
    Separating these is necessary to isolate the dynamics of interest and is an underdetermined data-analysis task.
    For the simplest case, with one triggered state and one reference state, a range of methods exists, but their assumptions, accuracy, and domains of applicability have not been systematically investigated.
    Here, we deconstruct and review the methods in current use, employing a unified notation. We make their assumptions explicit and benchmark each on a simulated dataset.
    A key point is that many of these methods do not estimate the fraction of all triggered molecules in the crystal, but instead estimate a latent parameter we call the extrapolation factor, which is related to but distinct from the so-called ``occupancy''.

    As our simulations make apparent, current methods do not always yield accurate results, in particular with noisy data and when the fraction of triggered protein is low.
    Inspired by previous work, we introduce Vacuum Level Matching, a method that grounds the estimate in the physical constraint that electron density is a nonnegative quantity: it cannot fall below a vacuum level.
    Operating voxel by voxel, Vacuum Level Matching replaces the manual choices of previous methods with a data-driven procedure and reports quality-control metrics that flag data unable to support a confident prediction.
    On the same simulated data, it outperforms state-of-the-art methods, and on two experimental datasets, rsEGFP2 and a multi-time-point photolyase series, it recovers known structural changes while flagging issues where the data are weak.
    Finally, we report an implementation of Vacuum Level Matching in the form of a small, reusable Python package.
\end{abstract}
\clearpage

\section{Introduction}

Time-resolved crystallography (TRX) has witnessed a surge of interest in recent years, as advances in X-ray sources and detectors have made it increasingly feasible to move from static structural snapshots to observing biological macromolecules in action.
By capturing metastable intermediates, populated on timescales from femtoseconds to seconds, researchers aim to visualize the precise atomic motions required for ligand binding, catalysis, and signal transduction \citep{khusainovTimeRevolutionMacromolecular2024,brandenAdvancesChallengesTimeresolved2021}.

The power of this approach is exemplified by recent studies.
Our own work used TRX to decipher the mechanism of DNA repair by the enzyme photolyase, allowing us to visualize photoexcitation, the subsequent breaking of thymine dimers, and product release \citep{christouTimeresolvedCrystallographyCaptures2023}.
In another case, TRX has been employed in conjunction with electric field stimulation to observe the mechanism of sodium ion conduction through the NaK2K channel \citep{leeDirectVisualizationElectricfieldstimulated2025}.

While TRX has largely been pursued by a small number of experts, the field is growing as the various enabling technologies mature.
For instance, while much of the research and development over the past decade has been driven by X-ray free-electron laser (XFEL) facilities, synchrotrons -- where beamtime is in greater supply -- are increasingly developing capable TRX-specific beamlines \citep{pearsonSerialSynchrotronCrystallography2020}.
Simultaneously, innovations in sample delivery for serial crystallography, in which diffraction from many microcrystals is merged into a single dataset, have lowered sample consumption requirements using both liquid jets and fixed targets, expanding the range of protein systems that can be studied \citep{barendsSerialFemtosecondCrystallography2022,banariAdvancingTimeresolvedStructural2025}.
Finally, the programs and data-analysis pipelines that facilitate TRX have matured into usable software packages \citep{grosse-kunstleveComputationalCrystallographyToolbox2002, whiteCrystFELSoftwareSuite2012, dezitterXtrapol8EnablesAutomatic2022, fadiniDenoisingIterativePhase2025}.
Consequently, there is hope that users who are not TRX experts will routinely use the technique to address new biological questions.

Substantial barriers remain, however.
One important challenge involves the analysis of mixtures of structural conformations in TRX data.
In a TRX experiment, one typically triggers a dynamical process using a light flash from an optical laser, rapid mixing with a ligand, the application of an external field, or another suitable perturbation.
This impulse initiates the reaction or motions of interest that can be followed crystallographically.
Unfortunately, with this approach it is almost always impossible to achieve a fully \emph{triggered} sample, in which all the proteins in the crystal are triggered and remain structurally synchronized during the dynamics of interest.
First, triggering usually activates only a small fraction of the proteins in a crystal, leaving many unperturbed molecules in the ground or \emph{reference} state.
Second, since triggered proteins progress through their reaction intermediates stochastically, they may not all adopt the same structural conformation at any given time.
Therefore, in general, a triggered crystal in a TRX experiment contains a mixture of protein conformations.
Separation of this mixture, so that one can solve the structure of each intermediate that contributes, is a central data-analysis task for any TRX practitioner.

The simplest corresponding model of the crystallographic signal asserts that the observed complex structure factors can be described through a linear combination,
\begin{equation}
    \complex{\strucmix}(t) = \combi_0(t) \complex{\strucdark} + \combi_1(t) \complex{\structarget} + \combi_2(t) \complex{F}_2 + \cdots
\end{equation}
where we have written the structure factors $\FT\{ \rho(\rr)\} = \complex{F}(\qq)$. Complex values are denoted by boldface, and we drop the argument of $\complex{F}$ to lighten the notation.

This linear model rests on several assumptions about the sample that are rarely stated explicitly.
First, it assumes that each lattice site is occupied by one conformation, chosen at random according to the conformations' occupancies in a way that is statistically independent of the neighboring lattice sites.
The diffraction observed at the location of the Bragg peaks is then given by the Fourier amplitude of the average unit-cell electron density. The structure factors of the individual conformations add coherently, weighted by their occupancies.
If, instead, the lattice sites are correlated such that they are segregated into separate domains or crystals, the diffraction signal will include terms that sum over intensities rather than be proportional to a single coherent sum over their complex structure factors.
Second, both the populations $\combi_i(t)$ and the structure factors $\complex{F}_i$ are assumed to be fixed at each time point, although every measurement averages over a finite time resolution, set by, e.g., pump--probe jitter and pulse or exposure durations, during which the populations continue to evolve.
Finally, if the structure factors are estimated from multiple crystals, for example from a serial crystallography experiment, the above formula assumes that each crystal is prepared identically, with the same populations of each individual conformation.
How well these assumptions hold depends less on the data analysis than on the experiment itself, through choices such as crystal size, pump fluence, and time resolution.
We examine the effect of occupancy variance across crystals in \autoref{sec:occupancy-variance}.
However, under these assumptions, which can be met experimentally, this model provides an effective framework for analyzing time-resolved data that will be adopted throughout this manuscript.

Under this model, the problem at hand is to recover the population amplitudes $\combi_i$ and the intermediate complex structure factors $\complex{F}_i$ given a set of amplitudes $\strucmix(t)$ measured at different time points $t$. Ideally, many time points are measured and the number of distinct observations $\strucmix(t)$ greatly exceeds the number of intermediates of interest.
The matrix whose columns are the $\complex\strucmix(t)$ then has far more columns than its rank, which equals the number of linearly independent intermediates.
In this ideal situation, low-rank methods can be used to infer the underlying structure factors and populations \citep{schmidtApplicationSingularValue2003}.
Unfortunately, in current practice, TRX time points are extremely expensive to obtain, and oversampling in time is rarely possible.
When the number of achievable time points is on the same order as or smaller than the number of structural intermediates of interest, a pragmatic approach is to choose time points $t$ for which -- excepting any remaining untriggered protein -- a single intermediate is overwhelmingly populated and the contribution of other intermediates is negligible.
Our focus in this paper will be exclusively on this more common data-limited regime.

If this condition is met and at a specific time $t$ a single intermediate is overwhelmingly populated, the observed structure factors will approximately be
\begin{equation}
    \label{eqn:structmix}
    \complex{\strucmix} \approx \combi_0 \complex{\strucdark} + \combi_1 \complex{\structarget}
\end{equation}
where $\combi_0 \complex{\strucdark}$ is the residual contribution from untriggered protein molecules that remain in the reference state, and $\combi_1 \complex{\structarget}$ corresponds to the signal from the time-evolving population, which is the signal of interest.
As we now discuss, isolating this signal is, unfortunately, an underdetermined problem.

Some quantities are known.
As stated above, the amplitudes $\strucmix(t)$ are known directly from the time-resolved measurement.
By obtaining a reference structure before the sample is triggered, one can determine the amplitudes $\strucdark$ and the corresponding phases of $\complex{\strucdark}$.
Assuming two states gives $\combi_0(t) + \combi_1(t) = 1$, leaving a single population parameter $\combi_1(t)$ -- commonly referred to as the \textit{occupancy} and denoted simply by $\occu$.
The remaining unknown quantities are thus the occupancy $\occu$, the phases of $\complex{\strucmix}$, and the complex structure factors $\complex{\structarget}$.

From \autoref{eqn:structmix}, it is apparent that with estimates of both $\occu$ and the phases of $\complex{\strucmix}$, it would be possible to solve for $\complex{\structarget}$.
Given an estimate for $\complex{\strucmix}$, the task of estimating an appropriate $\occu$ is commonly referred to as \emph{occupancy determination}. However, this turns out to be very difficult and is rarely done in practice; we show that most practitioners actually solve a simpler, but still effective, problem that we refer to as \emph{extrapolation factor determination}.
This extrapolation factor determination task is how most TRX experiments currently separate mixtures of states.
Because the problem is intrinsically underdetermined, any solution must introduce additional assumptions to determine $\occu$ and the phases of $\complex{\strucmix}$ -- or $\complex{\structarget}$ outright.

One recent approach sidesteps occupancy estimation altogether.
Candidate changes are enumerated along physics-informed reaction coordinates, and each candidate is scored by how well its calculated difference map reproduces the observed one \citep{fongRefinementAminoacidConformation}.
This applies only to a confined region, which must be specified in advance together with its reaction coordinates.
The more general route, and our focus here, is to estimate the extrapolation factor directly.
Methods for extrapolation factor estimation exist, each making pragmatic physical assumptions to resolve the inherent data ambiguities \citep{pearceMulticrystalMethodExtracting2017, coquelleChromophoreTwistingExcited2018, tripathiPHDependencePhotoactive2012}.
However, these assumptions and their domains of applicability have not been extensively assessed to date.
As a consequence, users have no basis on which to choose the method that best matches their situation, and little idea of what accuracy to expect.

Motivated by the lack of a comprehensive review and analysis of extrapolation factor determination methods, and interested in using one or more of those methods to solve our own TRX structures, we systematically investigated all methods published to date.
Here, we present our findings as well as a new method of our own.
We begin with a detailed presentation of the occupancy estimation problem, including the constant-phase approximation that underpins almost all current methods.
We then briefly review each method in use today, employing a unified notation and making its assumptions explicit.
Furthermore, we benchmark their performance on a simulated dataset and show how our method performs on two diverse, representative TRX datasets.

These results inspired us to develop a new method, which we call \textit{Vacuum Level Matching}, that grounds the estimation in the physical constraint that electron density cannot fall below the density of vacuum.
This method is an evolution of occupancy estimation techniques based on non-negativity that are already in use today \citep{schmidtPracticalConsiderationsAnalysis2023}.
Vacuum Level Matching operates at the level of individual voxels, replaces the manual fitting choices required by some existing methods with a data-driven cutoff, and self-reports quality-control metrics that allow it to flag cases where the data do not support a confident prediction.
We provide Vacuum Level Matching as a small, reusable software package so that it can be applied directly to new datasets.
These results should be useful to anyone analyzing TRX data.

\section{Structure factor extrapolation and the role of phases}
\label{sec:extrapolation-factor}

Above, we established that separating the triggered and reference structure-factor amplitudes is underdetermined. The natural approach would be to subtract some reference signal from the mixed state, leaving behind only the triggered signal of interest. Here, we review this idea in detail.

\subsection{Extrapolation with full phase information}
\label{sec:xtr-complex}

In the interest of clarity, we begin with the simplest case.
Suppose that, in addition to the amplitudes normally measured, we have accurate estimates of the phases and therefore the complex structure factors of both the reference state $\complex{\strucdark}$ and the mixed state $\complex{\strucmix}$.
The mixed-state structure factor is
\begin{equation}
    \complex{\strucmix} = (1 - \occu) \complex{\strucdark} + \occu \complex{\structarget},
    \label{eq:fmix-complex}
\end{equation}
which can be rearranged to give the triggered-state structure factor as a linear combination of the reference and the complex difference structure factors $\complexstrucdiff = \complex{\strucmix} - \complex{\strucdark}$,
\begin{equation}
    \complex{\structarget} = \complex{\strucdark} + \complexstrucdiff/\occu.
\end{equation}
Reading this as an extrapolation procedure, one recovers $\complex{\structarget}$ from the reference and the difference by scaling the latter by $1/\occu$.

If, through some suitable procedure, we could infer an \emph{extrapolation factor} $\extrapol$ such that $\extrapol \approx \occu$, then we could compute a complex \emph{extrapolated structure factor}

\begin{equation}
    \complex{\strucxtr} = \complex{\strucdark} + \xtrfacnse\,\complexstrucdiff
    \label{eq:xtr-complex}
\end{equation}
as the optimal, lowest-error estimate of $\complex{\structarget}$.
With access to the complete complex information, $\extrapol$ simply equals $\occu$.

\autoref{eq:fmix-complex} inherits the assumptions of the general linear model discussed in the introduction and adds one more: it treats the time-evolving intermediate $\complex{\structarget}$ as a single, pure state, although at a general time delay the system may be in a mixture of states.
Moreover, TRX experiments are often performed in a serial fashion, so the occupancy may vary across merged crystals, and where this variance is substantial it should be taken into account, as discussed in \autoref{sec:occupancy-variance}.

\subsection{Extrapolation with reference-state phases only}
\label{sec:xtr-darkphases}

While in the preceding section we assumed knowledge of the phases of $\complex{\strucmix}$, in practice these cannot be measured.
\autoref{eq:xtr-complex} is therefore not directly usable.
The pragmatic substitution that underpins almost all current methods is to replace the unknown phases of $\complexstrucdiff$ with the known phases of $\complex{\strucdark}$,
\begin{equation}
    \complexstrucdiff \;\to\; \strucdiffabs \, e^{i\phase_0},
    \label{eq:darkphase-substitution}
\end{equation}
where $\strucdiffabs = \strucmix - \strucdark$ is the isomorphous difference of the observed amplitudes, and $\phase_0$ is the reference-state phase.
Under this reference-phase approximation, \autoref{eq:xtr-complex} becomes a relation between amplitudes,
\begin{equation}
    \strucxtr = \strucdark + \xtrfacnse \, \strucdiffabs.
    \label{eq:xtr-definition}
\end{equation}
The errors and limitations of this scalar approximation to structure-factor extrapolation have been analyzed in detail by \citet{genickStructurefactorExtrapolation2007}.

How good is this approximation, and what is the optimal choice of $\extrapol$ in this setting?
The substitution \autoref{eq:darkphase-substitution} combines the measured amplitude difference with the reference-state phases. For reflections where the complex difference is small relative to the reference amplitude, $|\complexstrucdiff| \ll \strucdark$, the first-order approximation gives \citep{luzzatiTraitementStatistiqueErreurs1952,hendersonDifferenceFourierTechnique1971}
\begin{equation}
    \strucdiffabs \, e^{i\phase_0} \approx
    \frac{1}{2}\complexstrucdiff
    + \frac{1}{2}\strucdiff \, e^{i 2\phase_0 - i\phase_\Delta},
    \label{eq:fxtr-complex}
\end{equation}
where $\phase_\Delta$ is the true phase of $\complexstrucdiff$ (see \autoref{sec:occupancy-variance} for the derivation with crystal-to-crystal variance).
The isomorphous difference therefore contains the true difference at half-scale, plus the complex conjugate term carrying phase $e^{i 2\phase_0 - i\phase_\Delta}$.
To interpret this decomposition, we make the additional approximation that the conjugate contribution is diffuse across the unit cell and negligible in regions containing the structural changes.
This behavior is not guaranteed by the decomposition itself: correlations between the reference and difference phases can leave a structured residual.

Substituting \autoref{eq:fxtr-complex} into \autoref{eq:xtr-definition}, together with the definition of $\strucdiff$, yields
\begin{equation}
    \strucxtr \, e^{i\phase_0}
    \approx \frac{\xtrfacnse\,\occu}{2}\complex{\structarget}
    + \left(1 - \frac{\xtrfacnse\,\occu}{2}\right)\complex{\strucdark}
    + \frac{\xtrfacnse}{2}\,\complexstrucdiff^* e^{i 2\phase_0}.
    \label{eq:xtr-noise}
\end{equation}
The final term is the conjugate contribution discussed above.
Under the approximation that this contribution is diffuse and negligible in regions containing the structural changes, $\extrapol = \occu/2$ restores the coherent difference signal to its full amplitude.
We use ``optimal'' in this sense of recovering the amplitude of the local structural features, rather than necessarily minimizing the total reconstruction error across the unit cell.
The latter criterion also penalizes the amplified conjugate contribution and can favor a different extrapolation factor.

\subsection{The effect of structure factor weighting}
\label{sec:xtr-sf-weighting}

In current TRX analysis, it is standard to apply scalar weights $\weightfousf$ to the difference structure factors to mitigate outliers and improve the signal-to-noise ratio in the resulting map \citep{ursbyImprovedEstimationStructureFactor1997, renMolecularMovie182001, dezitterXtrapol8EnablesAutomatic2022}.
Per reflection, $\weightfousf$ enters as an additional prefactor on $\strucdiff$, usually depending on measurement error and the absolute values of both $\strucdark$ and $\strucmix$, so \autoref{eq:xtr-noise} becomes
\begin{equation}
    \strucxtr \, e^{i\phase_0}
    = \frac{\xtrfacnse\,\weightfousf\,\occu}{2}\complex{\structarget}
    + \left(1 - \frac{\xtrfacnse\,\weightfousf\,\occu}{2}\right)\complex{\strucdark}
    + \frac{\xtrfacnse\,\weightfousf}{2}\,\complexstrucdiff^* e^{i 2\phase_0}.
    \label{eq:xtr-noise2}
\end{equation}
We keep the factor $1/2$ separate from $\weightfousf$ here to distinguish the effect of the reference-phase approximation from that of the weighting scheme, whereas in \autoref{eq:xtr-general} both are combined.
The optimal extrapolation factor is now
\begin{equation}
    \extrapol = \frac{1}{2}\,\occu\,\weightfousf.
    \label{eq:xtr-sf-weighting-optimal}
\end{equation}
This complicates the direct determination of occupancy in two ways.
First, the optimal extrapolation factor is now tied to the specific weighting scheme applied.
Second, the relationship between $\extrapol$ and $\occu$ is no longer a simple proportionality, since $\weightfousf$ varies across reciprocal space.

\subsection{Phase modification}
\label{sec:xtr-phases}
As discussed above, the true structure-factor amplitudes $\strucdiff$ are unknown, and only the differences of the absolute amplitudes $\strucdiffabs$ are measured.
Real-space refinement techniques that impose constraints such as TV denoising \citep{fadiniDenoisingIterativePhase2025} necessarily modify both amplitudes and phases.
Moreover, real-space operations do not necessarily preserve crystallographic symmetries, possibly introducing nonzero structure factors at symmetry-forbidden reflections in either $\strucmix$ or $\strucdark$.
These effects compound the uncertainty in the relationship between estimated structure factors and the true $\complexstrucdiff$, further obscuring the connection between $\extrapol$ and $\occu$.

One remedy is to iteratively project between real-space constraints (such as TV denoising) and the known structure factors $\strucmix$ and $\complex{\strucdark}$ \citep{fadiniDenoisingIterativePhase2025}. This can improve the amplitude and phase estimate for the difference structure factors, but to an unknown extent.
As a result, one effectively obtains $\extrapol = \occu\weightfouip$ with a complex-valued weighting parameter.
Although iterative phase refinement may improve phase estimates, its effects on both amplitudes and phases can increase uncertainty in the relationship between $\extrapol$ and $\occu$.

\subsection{Relationship between the extrapolation factor and occupancy}
\label{sec:xtr-occupancy-relationship}
The preceding sections show that the optimal extrapolation factor depends not only on the underlying physical occupancy, but also on the phase accuracy and post-processing applied to the difference map.
We therefore recommend treating the extrapolation factor and the physical occupancy as related but distinct quantities, and stating the processing-pipeline specifics alongside any reported extrapolation factor.
For this reason, we prefer the term \emph{extrapolation factor determination} over the more common \emph{occupancy determination}, since what these methods recover is the extrapolation factor, and only under the idealized conditions of \autoref{sec:xtr-complex} does it coincide with the physical occupancy.

In general, then, the extrapolated structure factor can be written as

\begin{equation}
    \complex{\strucxtr}
    = \left(\xtrfacnse\occu\weightfougeneral\right)\complex{\structarget}
    + \left(1 - \xtrfacnse\occu\weightfougeneral\right)\complex{\strucdark}
    + \left(\xtrfacnse\occu\weightfougeneral\right)\complex{N}_{\strucdiff},
    \label{eq:xtr-general}
\end{equation}
where $\weightfougeneral$ acts as a general Fourier weight that captures the applied data modifications, including the $1/2$ term from the reference phase approximation.
The $\complex{N}_{\strucdiff}$ term captures the noise and error in the difference map, both from a potential reference-phase approximation and from any other errors due to incorrect structure factors or imperfect phase refinement.

Unlike in \autoref{eq:xtr-noise} and \autoref{eq:xtr-noise2}, here we factor $\occu$ out of this term and absorb a compensating $1/\occu$ into the definition of $\complex{N}_{\strucdiff}$, so that the real-space form in \autoref{eq:vm-realspace} reads more compactly.
It is thus crucial to know and report the effect of the difference map calculation pipeline to make a substantive claim about the relationship between $\extrapol$ and $\occu$.

\section{Estimating the extrapolation factor}
\label{sec:xtr-techniques}

Because occupancy determination is underdetermined, every method for estimating occupancy must introduce additional assumptions to obtain a solution.
These assumptions define each method's domain of applicability and its failure modes.
In this section, we survey the methods published to date, presenting each on its own terms before benchmarking them empirically in \autoref{sec:comparison}.
Throughout this discussion, we assume that besides complex reference structure factors $\complex\strucdark$, one has some estimate of the difference-map phases $\complex \strucdiff$, so that we can determine the real-space counterparts $\densdark$ and $\densdiff$ as well as similar objects.
Although we write these densities as continuous functions of the real-space coordinate $\rr$, in practice each is sampled on a discrete grid, and we refer to an individual grid sample as a voxel.
We make the dependence on $\rr$ explicit only when referring to values at specific points, such as $\densdiffr$, and leave it implicit otherwise.

\begin{figure}
    \centering
    \includegraphics[width=0.5\linewidth]{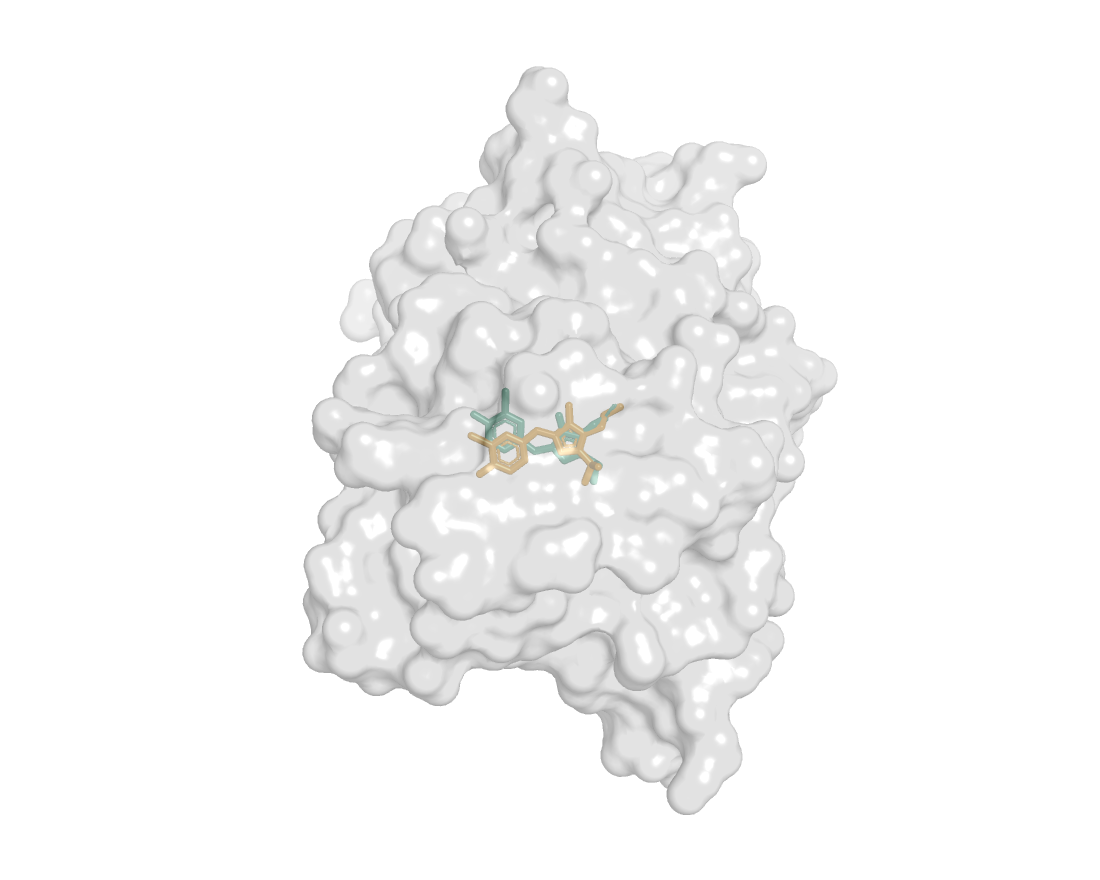}
    \caption{\textbf{Illustration of modeled change.} Here we show rsEGFP2 based on PDB entry 8A6G changing from the \textit{trans} reference state (green) to the \textit{cis} triggered state (orange).}
    \label{fig:example_vis}
\end{figure}

We begin with some approaches that have appeared in early TRX studies (\autoref{sec:classical}), then turn to the methods most widely used in recent work: PanDDA (\autoref{sec:pandda}), Difference Map Matching (\autoref{sec:x8}) and the Negative Sum Explosion (\autoref{sec:nse}).
The latter is of particular interest to us, because it rests on a single, transparent physical constraint.
In \autoref{sec:vacuum-matching} we build on it directly, introducing \emph{Vacuum Level Matching} as our own modification: it keeps the same negative-density principle but resolves the estimate voxel by voxel rather than aggregating over the whole map.
In the next section, we test our method as well as the other more recent methods on a simulated dataset.
From these simulations, we take one run and use it as an example to illustrate each method.
To achieve a realistic simulation, we computed signals based on the \textit{cis} and \textit{trans} chromophore structures of rsEGFP2 determined by a previous time-resolved serial femtosecond crystallography (TR-SFX) experiment \citep{fadiniSerialFemtosecondCrystallography2023}, using a simulated occupancy of $\occusim = 0.4$ with the reference-phase approximation (target extrapolation factor $\extrapol = 0.2$).
Throughout this section, we will show a representative example with medium noise.
Further details about the setup of the simulation can be found in \autoref{sec:sim-methods}.
The model and the change can be seen in \autoref{fig:example_vis}.

\subsection{Model- and volume-based approaches}
\label{sec:classical}

Three approaches predate the methods we will benchmark in detail and are included below for completeness.
Each has been applied successfully in foundational TRX studies, and each illustrates a distinct way of grounding the extrapolation factor estimate.
We do not include them in our quantitative comparison.
B-factor matching and distance analysis require a model for the triggered state that is refined at multiple candidate extrapolation factors and so cannot be evaluated on the same purely map-based footing as the other methods, while integrated difference electron density requires manual isolation of an integration volume that does not generalize across our test datasets.

\subsubsection{Integrated difference electron density}
\label{sec:integrated-ded}
One approach to estimating the extrapolation factor is to identify a region of density in $\densdiff$ corresponding to a complete, well-delimited density feature, such as a ligand, in $\densmix$.
The theoretical electron count $N_\text{e}$ of this species is known from its chemical composition.
With the maps placed on an absolute scale, the difference density is then integrated over this region $V$.
The extrapolation factor is chosen such that the extrapolated difference density $\xtrfacnse\,\densdiff$ integrates to the full electron count $N_\text{e}$, that is,
\begin{equation}
    \extrapol
    = \frac{\bigl|\int_V \densdiff \, \mathrm{d}\rr\bigr|}{N_\text{e}},
    \label{eq:integrated-ded}
\end{equation}
so that the extrapolated map contains the complete species.
No explicit correction for the reference-state phases is required, since the loss of half the coherent signal (\autoref{sec:xtr-darkphases}) is already reflected in the measured integral.
Notable applications include the determination of CO photolysis yields in myoglobin \citep{srajerPhotolysisCarbonMonoxide1996} and the estimation of photoinitiation levels in photoactive yellow protein \citep{schmidtProteinKineticsStructures2004}.

The principal difficulty lies in defining precise volume boundaries.
This requires a region free of overlapping signal, which is difficult to identify reliably, especially at low occupancy or high noise.
Even when the difference features are large and well defined, the boundaries must, at present, be drawn by hand and rely on human judgment.

\subsubsection{B-factor matching}
This method exploits the strong correlation between crystallographic B-factors and the extrapolation factor during structural refinement.
It assumes that the B-factors, as a measure of local disorder, of atoms that move after the trigger do not differ drastically from those of their neighboring, unchanged atoms.
Since both occupancy and B-factors scale the atomic scattering contribution, an error in the assigned occupancy is often absorbed by an artificial shift in the B-factors.
To identify the correct extrapolation factor, a model is refined against the data at various extrapolation levels.
The optimal extrapolation factor is determined by finding the point where the B-factors of atoms that move after perturbation match those of the surrounding, static protein environment.

This approach was successfully employed by \citet{nangoThreedimensionalMovieStructural2016} to determine the structural changes in bacteriorhodopsin.
In their implementation, they acknowledged that simultaneous refinement of B-factors and extrapolation factors was unstable at their resolution due to parameter coupling.
Instead, they refined the coordinates and extrapolation factors of the triggered-state conformers while keeping B-factors constrained.
To derive a final consensus extrapolation factor for each time point, they calculated a weighted average based on the grouped residues that exhibited the largest structural displacements.

\subsubsection{Distance analysis}
\label{sec:distance-analysis}
The core principle behind distance analysis was first articulated by \citet{genickStructureProteinPhotocycle1997}.
An incorrect assumed extrapolation factor during refinement produces structural artifacts in the refined model.
When the assumed $\xtrfacnse$ is too low, extrapolation does not fully remove the reference-state contribution, and the refined atomic positions remain biased toward the reference state.
Once the assumed $\xtrfacnse$ meets or exceeds the true value, the positions are expected to stabilize.
This observation was turned into a systematic procedure in Xtrapol8 \citep{dezitterXtrapol8EnablesAutomatic2022}, where it is referred to as ``distance analysis''.

In this workflow, models are refined against extrapolated structure factors generated over a range of assumed extrapolation factors.
Changes in interatomic distances relative to the reference model are then fitted with sigmoidal functions of $\xtrfacnse$.
For each accepted fit, the value of $\xtrfacnse$ at which the distance change reaches 99~\% of its fitted plateau is extracted.
These pairwise estimates are aggregated to obtain a consensus estimate.
The criterion identifies the regime in which further extrapolation produces little additional structural displacement.
Unlike the other methods presented here, distance analysis requires refining the model in both reciprocal and real space at every tested extrapolation factor.

\subsection{Comparing local and global correlations}
\label{sec:pandda}
\begin{figure}[htbp]
    \centering
    \includegraphics[width=0.6\linewidth]{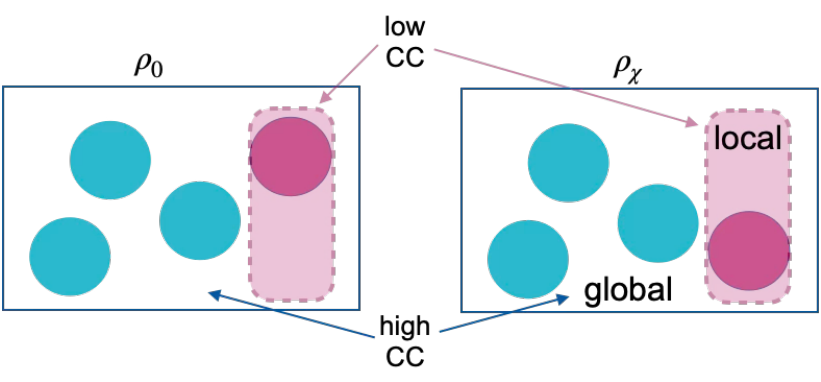}
    \caption{
        \textbf{Motivation for the PanDDA method.} We show a model molecule with four atoms in the reference state $\densdark$ and extrapolated state $\densxtr$.
        Three atoms (teal) stay in place, whereas a localized structural change is represented by the fourth (crimson).
        The method calculates the global and local cross-correlations (CCs) over a region of interest between $\densdark$ and $\densxtr$, iterating over many different values of $\extrapol$.
        At the optimal extrapolation factor, the global CC ought to be high but the CC in the local region (dashed line) ought to be low.
        The method selects as optimal the value of $\extrapol$ for which the spread between the two CCs is highest.
    }
    \label{fig:pandda-idea}
\end{figure}

\begin{figure}[htbp]
    \centering
    \includegraphics[width=0.6\linewidth]{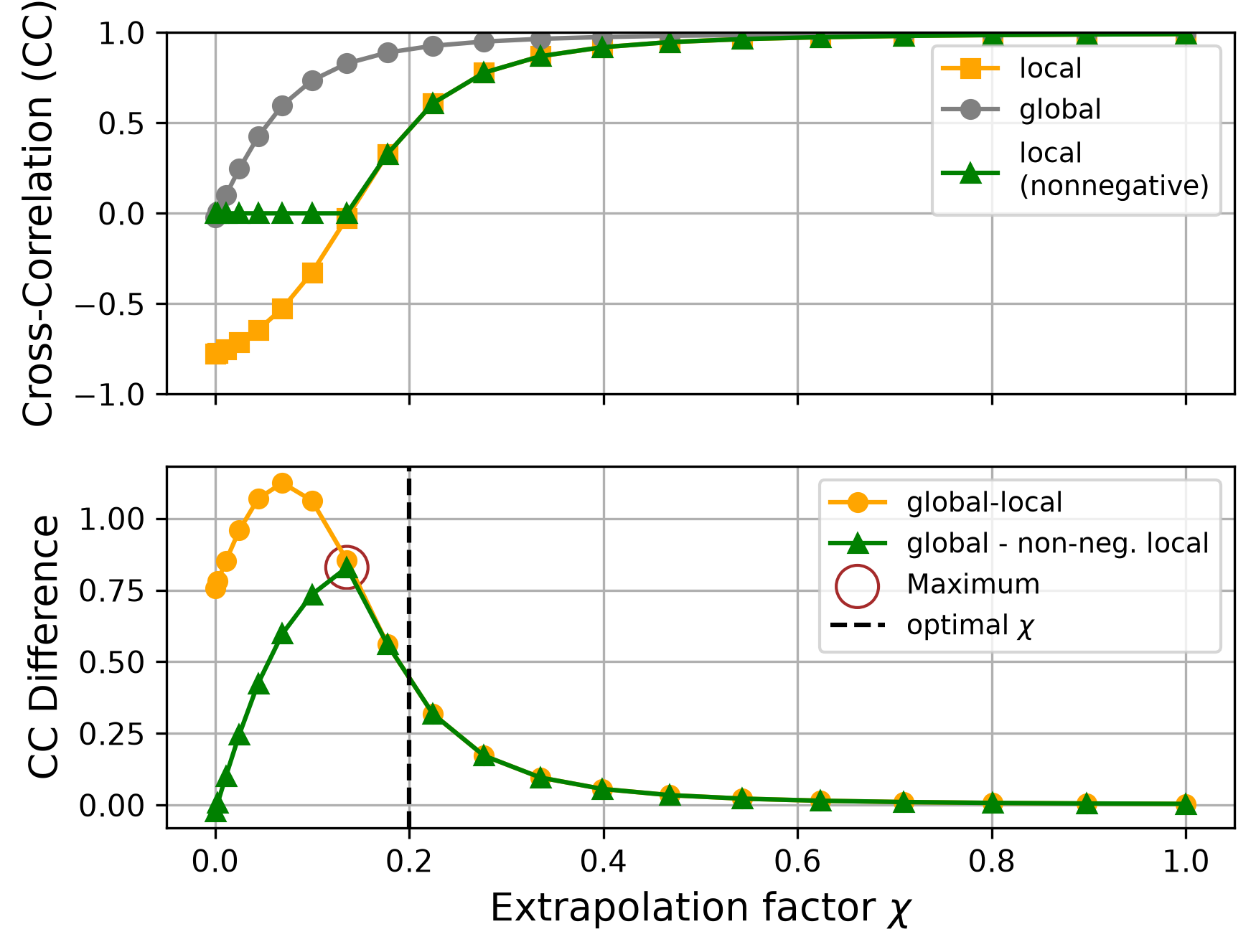}
    \caption{
        \textbf{Showcase example of the `PanDDA method' in action.}
        The top panel shows the global (grey), local (orange) and nonnegative local (green) correlations as a function of $\extrapol$, where only voxels with $|\densdiffr| > 3 \rmsddiff$ are included.
        The bottom panel shows the difference between the global correlation and either the local (orange) or the nonnegative local (green) correlation.
        The results are based on a single simulated dataset with an expected $\extrapol = 0.2$ (modeled with $\occusim = 0.4$ and the reference-phase approximation) at an SNR of 15, as described in \autoref{sec:sim-methods}.
    }
    \label{fig:example_pandda}
\end{figure}

PanDDA \citep{pearceMulticrystalMethodExtracting2017} is a software tool widely used in fragment-based drug discovery, where its purpose is to detect weakly bound ligands in crystallographic datasets.
Within this tool, a routine determines the \emph{background density component}, which is operationally equivalent to $1 - \extrapol$.
While neither PanDDA nor its routine for determining the background component was designed for time-resolved data, the same routine -- informally referred to as the `PanDDA method' -- has subsequently been applied in TRX studies \citep{fadiniSerialFemtosecondCrystallography2023}.

The central idea is to maximize the contrast of an extrapolated map in real space, such that the density difference of a changing region is as large as possible.
The routine begins by identifying a \emph{local} region of interest where changes occur.
The entire unit cell is treated as the \emph{global} region.
Then, for a range of $\extrapol$, the cross-correlation of the extrapolated map $\densxtr = \densdark + \xtrfacnse\,\densdiff$ with the reference map $\densdark$ is computed for both the local and global regions.
This local region of interest can be determined in various ways.
A simple option is to threshold the absolute difference map, selecting all voxels whose value exceeds three times the difference-map \rmsd{} (i.e.\ $|\densdiffr| \geq 3\,\rmsddiff$).

The behavior of the two cross-correlations, as $\extrapol$ is varied, is intuitive.
At small $\extrapol$, $\densxtr$ is dominated by $\densdiff$, and both cross-correlations approach zero.
At large $\extrapol$, $\densxtr$ closely resembles $\densdark$, and both global and local cross-correlations are near unity.
Because the difference map is `most different' in the local region, the local cross-correlation rises only at larger $\extrapol$ than the global one (see \autoref{fig:pandda-idea}).

The maximum of the difference between the global and local cross-correlations was proposed as the estimator of the correct extrapolation factor \citep{pearceMulticrystalMethodExtracting2017}.
An application of this method can be seen in \autoref{fig:example_pandda}.

\paragraph{Adaptation to time-resolved data}
A complication arises when applying the routine to TRX rather than ligand-binding data.
In the ligand-binding setting for which PanDDA was designed, the local difference signal is predominantly positive, since ligand atoms occupy previously empty space.
In the TRX setting, by contrast, the difference signal contains strong negative features wherever atoms have vacated their reference-state positions.
As a consequence, the local cross-correlation is not merely absent but negative at small $\extrapol$, and the trajectory of the local-versus-global difference is distorted relative to the regime in which the heuristic was originally calibrated.
A simple modification is to restrict the cross-correlation to nonnegative voxel values, suppressing the inverted starting point.
We refer to this as the \emph{nonnegative PanDDA} variant.

The PanDDA routine has the appeal of operating directly on cross-correlations between maps that are easy to compute.
Two features of the method are worth noting in comparison with the others.
The maximum-difference criterion is a heuristic.
The original PanDDA paper reports that this criterion correlates with the final occupancy estimate, with a nontrivial variance \citep{pearceMulticrystalMethodExtracting2017}.
This relationship is empirical and it does not follow from an independent physical constraint on the extrapolated map.
Its accuracy on data outside the original fragment-screening domain has not been systematically characterized.

Moreover, the result is sensitive to the local region mask.
Different filtering choices produce different masks and therefore different local cross-correlations, which can shift the position of the maximum and thus the estimated $\extrapol$.
What can be said is that the optimal extrapolation factor lies within the range where the global-minus-local difference is appreciably nonzero: when the global and local cross-correlations are similar, this indicates either under-extrapolation (both high) or over-extrapolation (both low).
The reported maximum has no quantifiable guarantee of being a better estimate than any other point within that range.

We tested both the standard and nonnegative variants but show only the nonnegative version in \autoref{sec:comparison}. The standard variant, as described, has a tendency to over-extrapolate due to negative correlations in the local region.

\subsection{Difference Map Matching}
\label{sec:x8}

\begin{figure}[htbp]
    \centering
    \includegraphics[width=0.6\linewidth]{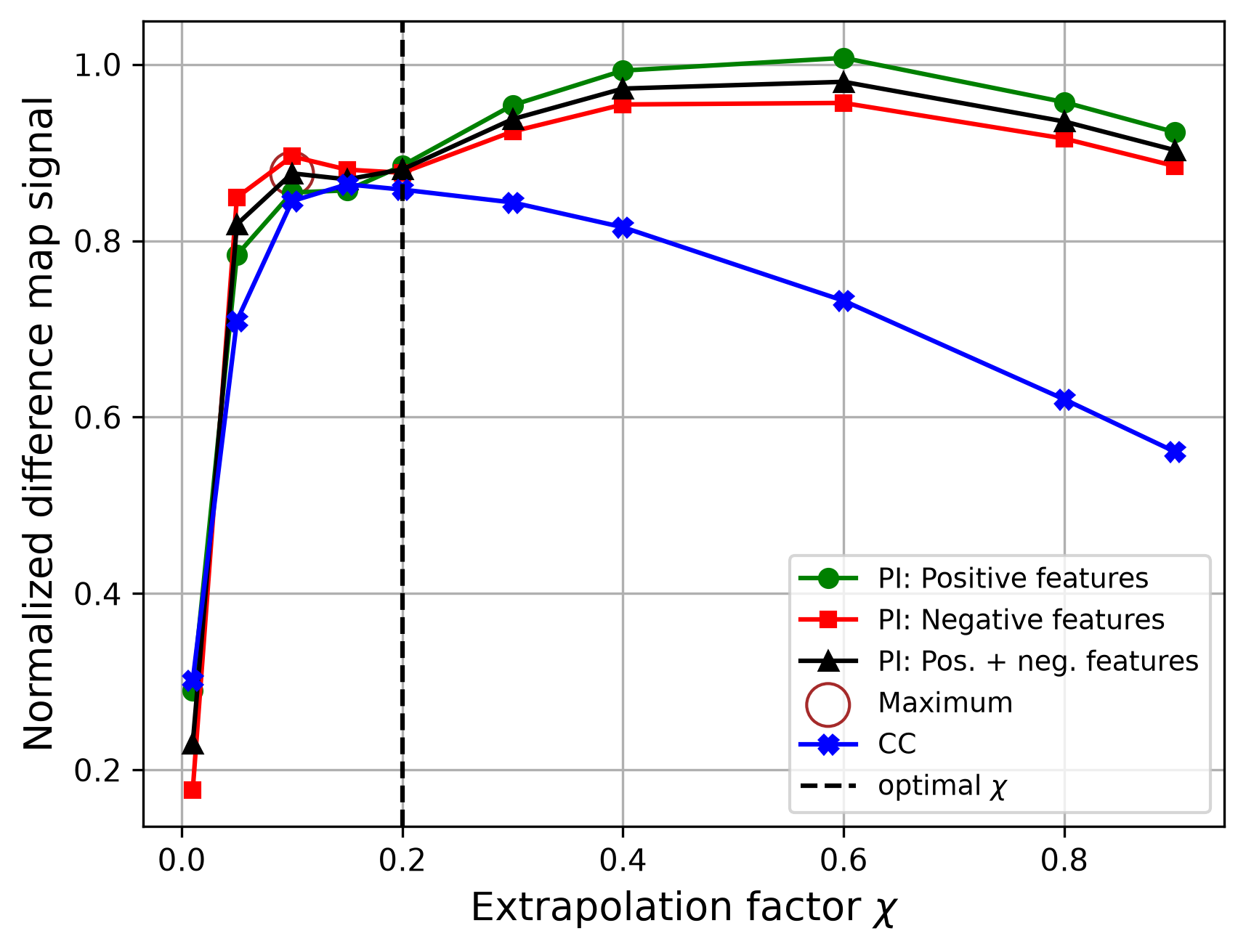}
    \caption{
        \textbf{Showcase example of `Difference Map Matching' in action.}
        Difference Map Matching applied to the simulated rsEGFP2 dataset (target $\extrapol = 0.2$).
        Curves show the normalized difference-map signal as a function of $\extrapol$ for the Peak Integration (PI) method, evaluated over positive, negative, and combined features, together with the cross-correlation (CC) method.
        The Peak Integration ratio also rises towards large $\extrapol$ (under-extrapolation), but this spurious maximum is excluded by requiring that the extrapolated and observed difference maps agree in sign at the largest absolute voxels of $\densdifffofo$ (\autoref{sec:x8}).
        For this dataset, the Peak Integration method (all features) gives $\extrapol = 0.10$ and the CC method $\extrapol = 0.15$.
    }
    \label{fig:example_x8}
\end{figure}

The Difference Map Matching approach was first introduced by \citet{coquelleChromophoreTwistingExcited2018}, and a revised version is implemented in the software package Xtrapol8 \citep{dezitterXtrapol8EnablesAutomatic2022}.
For each candidate extrapolation factor, the method compares the observed and extrapolated difference maps after processing and normalization.
The optimal extrapolation factor is identified as the value that yields the closest agreement between the two maps.

In Xtrapol8, there are two flavors of this technique, which we refer to as the `Peak Integration method' and the `CC method' based on their respective approaches to discovering signal.
Both have been applied in several recent TRX studies \citep{hadjidemetriouTimeresolvedSerialFemtosecond2022, christouTimeresolvedCrystallographyCaptures2023, rios-santacruzIntegratedStructuralDynamics2026}.

At the heart of this method lie two quantities:
the difference between a given extrapolated structure and the calculated reference state, i.e.\ $\densdiffxtr = \weightxtr({\densxtr} - {\densdark^\text{calc}} )$, and the experimental `$F_o-F_o$' map, i.e.\ the observed differences $\densdifffofo=\weightobs(\densmix - \densdark)$,
where $W_i$ represents the linear post-processing applied to each difference density map $i$.
The goal of both types of difference map matching is to find the $\extrapol$ for which $\densdiffxtr$ best approximates $\densdifffofo$.
Both methods quantify the differences between $\densdiffxtr$ and $\densdifffofo$.
To see what differences are found, we rewrite $\densdiffxtr$ as a function of $\densdifffofo$,

\begin{align}
    \label{eqn:diffmap-matching}
    \densdiffxtr
    = \xtrfacnse{\densdifffofo}
    + \xtrfacnse \deltadiff
    + \deltadark,
\end{align}
where $\deltadark = \weightxtr(\densdark^\text{obs} - \densdark^\text{calc})$ captures the mismatch between the observed and calculated reference maps -- the conventional `$F_o-F_c$' map of the reference state.
The expression $\deltadiff = (\weightxtr -\weightobs)(\densmix - \densdark)$ absorbs the differences due to the separate `post-processing' steps.
For fixed post-processing weights, $\xtrfacnse\deltadiff$ grows with the extrapolation, whereas $\deltadark$ is independent of it; candidate-dependent post-processing can introduce additional variation in both mismatch terms.
Both methods rely on these mismatch terms being non-negligible, since it is their $\extrapol$-dependent contribution relative to $\xtrfacnse\densdifffofo$ that provides the contrast between candidate extrapolation factors.

\paragraph{Peak Integration Method.}

This method tests whether the extrapolated difference map reproduces the observed difference signal well by comparing the locations where that signal is most prominent.
First, a subset of prominent voxels $\rr_\text{signal}$ is selected from the observed difference map $\densdifffofo$ by choosing the largest peaks, that is, regions of large absolute values ($>3.5 \rmsddiff$ by default).
Both $\densdifffofo$ and $\densdiffxtr$ are then normalized by their respective \rmsd{}.
This step is essential: on an absolute scale, the integrated density at $\rr_\text{signal}$ grows linearly even under over-extrapolation, so the raw sum is uninformative.
Because over-extrapolation raises the density at all voxels, not only at $\rr_\text{signal}$, the normalized integral falls on either side of the optimum.
Next, $\densdiffxtr$ is integrated over this fixed mask, separately for positive and negative peaks (assigned by the sign of $\densdiffxtr$) and for both combined.
The resulting score for each channel ($\pm$) is
\begin{equation}
    \text{S}(\pm)
    = \frac{\sum_{\rr_\text{signal}}
        \bigl|\densdiffxtr(\rr_\text{signal}, \pm)\bigr|}
    {\sum_{\rr_\text{signal}}
        \bigl|\densdifffofo(\rr_\text{signal}, \pm)\bigr|}.
\end{equation}
All three channels $\pm$ (positive only, negative only, and combined) are displayed, but the optimal $\extrapol$ is chosen from the combined (positive-plus-negative) absolute sum.

By construction, most density in $\densdifffofo$ is concentrated at $\rr_\text{signal}$, whereas the disagreements in difference map processing $\deltadiff\propto\weightxtr-\weightobs$ and the disagreement between the observed and calculated reference maps $\deltadark\propto\densdark^\text{obs} - \densdark^\text{calc}$ are distributed across the entire map.
This metric looks for the extrapolated difference map with the highest concentration at $\rr_\text{signal}$.
In both the over-extrapolation and under-extrapolation regimes,
relatively more density is expected to accumulate outside of $\rr_\text{signal}$ compared to inside.
Because the sign channels are assigned from $\densdiffxtr$ alone, without reference to $\densdifffofo$, a spurious second maximum can appear, typically at large $\extrapol$, where the in-mask density has inverted in character (\autoref{fig:example_x8}).
To exclude it, we check that $\densdiffxtr$ has the same sign as $\densdifffofo$ at the voxels of largest absolute value in $\densdifffofo$, and accept a maximum only if this holds.

\paragraph{Cross-Correlation Method.}
The second routine compares the extrapolated difference map $\densdiffxtr$ directly to the observed difference map $\densdifffofo$ through their Pearson cross-correlation.
The correlation measures how closely the mean-centered spatial patterns of the two maps agree; the role of the mismatch terms is derived in \autoref{sec:diffmap-matching-derivation}.
Its global scope means it does not separate signal peaks from noise peaks.
Noise peaks therefore also contribute to the correlation, so stochastic differences in the solvent or in other regions of little change can bias it even though they carry no structural information.

The Peak Integration method works best when the conventional difference map contains a small number of well-isolated, prominent peaks, because the mismatch terms are then spread over the rest of the map and contrast sharply with the signal at $\rr_\text{signal}$.
It struggles when the density is diffuse or when signal and noise peaks have comparable prominence.
The CC method aggregates over the full map and is not sensitive to peak-selection choices, but the most important information is contained in the extreme values, which the Peak Integration method captures.

When the two sub-methods disagree, as they do in our example (\autoref{fig:example_x8}), there is no principled basis for choosing between them, because neither quantifies its confidence.
A further practical cost is that each sampled $\extrapol$ requires, in the present form, separate post-processing for each $\densdiffxtr$, which makes fine sampling expensive compared to the other methods surveyed.

We evaluate both the Peak Integration and CC methods in \autoref{sec:comparison}.

\subsection{Negative Sum Explosion}
\label{sec:nse}

\begin{figure}[htbp]
    \centering
    \includegraphics[width=\linewidth]{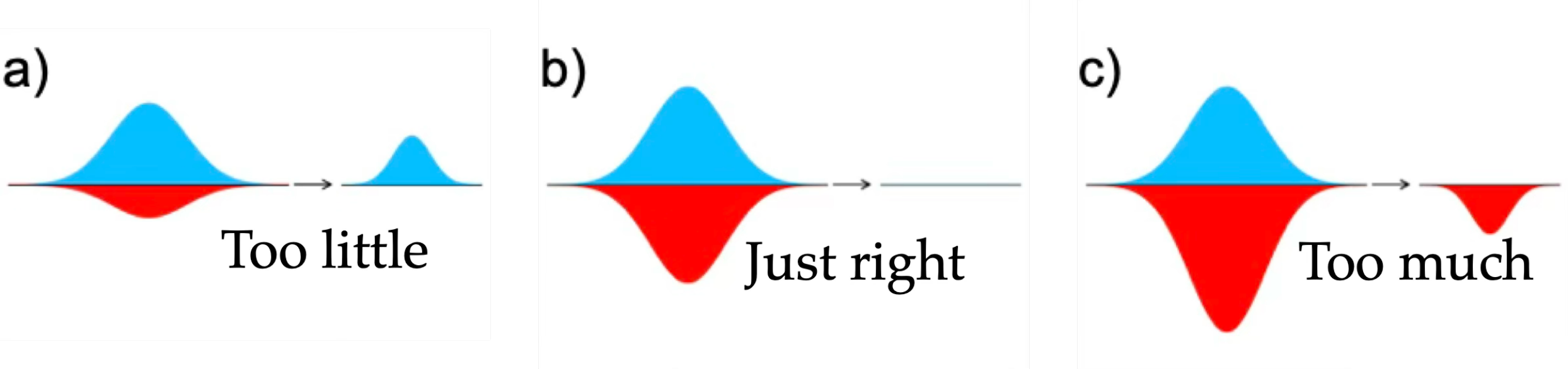}
    \caption{
        \textbf{Illustration of the principle behind NSE}.
        This figure (inspired by Fig. 8 from \citet{schmidtPracticalConsiderationsAnalysis2023}) shows three schematic additions of a positive reference density (blue) and a negative difference density (red) on the left side of the arrow and the resulting extrapolated density on the right side of the arrow.
        a) shows an example of under-extrapolation (``too little''), where extrapolated density remains positive.
        b) shows the correct extrapolation (``just right''), where extrapolated density is close to zero.
        c) shows an example of over-extrapolation (``too much''), where extrapolated density becomes negative.
    }
    \label{fig:nse-idea}
\end{figure}

Next, we discuss a technique first used in \citet{tripathiPHDependencePhotoactive2012} and further refined in \citet{schmidtPracticalConsiderationsAnalysis2023}. Because, to the best of our knowledge, this technique has not yet been given a name, we refer to it as the `Negative Sum Explosion' method (NSE).
The NSE has been used in several recent publications \citep{pandeyTimeresolvedSerialFemtosecond2020, baxterPowerDensityTitration2024,maestre-reynaCapturingStructuralIntermediates2025}.
It rests on a single physical constraint, namely, that any electron density map is expected to be predominantly nonnegative, with only minor negative signal arising from experimental error and Fourier truncation.
By contrast, over-extrapolation ($\extrapol<\trueextrapol$) yields negative density in regions of the extrapolated map $\densxtrr = \densdarkr + \xtrfacnse\densdiffr$ where $\densdiffr$ is negative (compare \subfigref{fig:nse-idea}{c}).
Under-extrapolation ($\extrapol>\trueextrapol$) means that negative density in the difference map is more than fully compensated by positive density in the reference map, so the level of negative density in $\densxtrr$ stays low and remains so even if $\xtrfacnse$ is increased slightly (compare \subfigref{fig:nse-idea}{a}).
Therefore, one way to estimate the extrapolation factor is to identify the point at which large amounts of negative density start to accumulate in $\densxtrr$ (i.e.\ \subfigref{fig:nse-idea}{b}).

Concretely, one computes $\densxtrr$ for a range of $\extrapol$ values and plots the sum over all negative voxels against $\xtrfacnse$.
In the under-extrapolation regime ($\extrapol$ is large and $\xtrfacnse$ is small), most voxels satisfy $\densdarkr > -\xtrfacnse\densdiffr$, so $\densxtrr$ remains positive and the negative sum stays approximately flat.
In the over-extrapolation regime ($\extrapol$ is small and $\xtrfacnse$ is large), most voxels with negative $\densdiffr$ satisfy $\densdarkr < -\xtrfacnse\densdiffr$ and the negative sum grows linearly with slope $\sum_{\rr:\,\densdiffr<0}\densdiffr$.
One therefore expects two approximately linear regimes separated by a short transition.
A linear regression is fitted to each regime, and the extrapolation factor is estimated as their intersection.

Two details are noteworthy:
\begin{enumerate}

    \item \textbf{The zero-frequency term}. An important point, not mentioned in the literature on this method, is the correction for the missing zero-frequency term $F_{000}$. This ``DC component'' is the constant, additive density contribution to every voxel in the map.
    Structure factors are conventionally reported without $F_{000}$, so the reconstructed real-space map has a zero mean rather than the vacuum-referenced mean of the true electron density.
    This artificially lowers $\densdarkr$ by a global constant, displacing the apparent zero-crossing of $\densxtrr$ and causing a systematic underestimation of $\xtrfacnse$.
    Tools exist, however, that estimate $F_{000}$ given the atomic model \citep{langProteinStructuralEnsembles2014}, including both the protein and solvent contributions.
    Applying them before analysis avoids this offset.

    \item \textbf{Focus on regions of interest}.
    Some $\densdiffr$ values are negative due to artifacts (noise, systematic error) while others are negative due to real structural changes. The NSE relies on the latter dominating the former.
    However, by focusing on regions of interest, we can mitigate the influence of artifacts, since structural changes are generally localized in space \citep{schmidtPracticalConsiderationsAnalysis2023}.

\end{enumerate}

\begin{figure}[htb]
    \centering
    \includegraphics[width=0.7\linewidth]{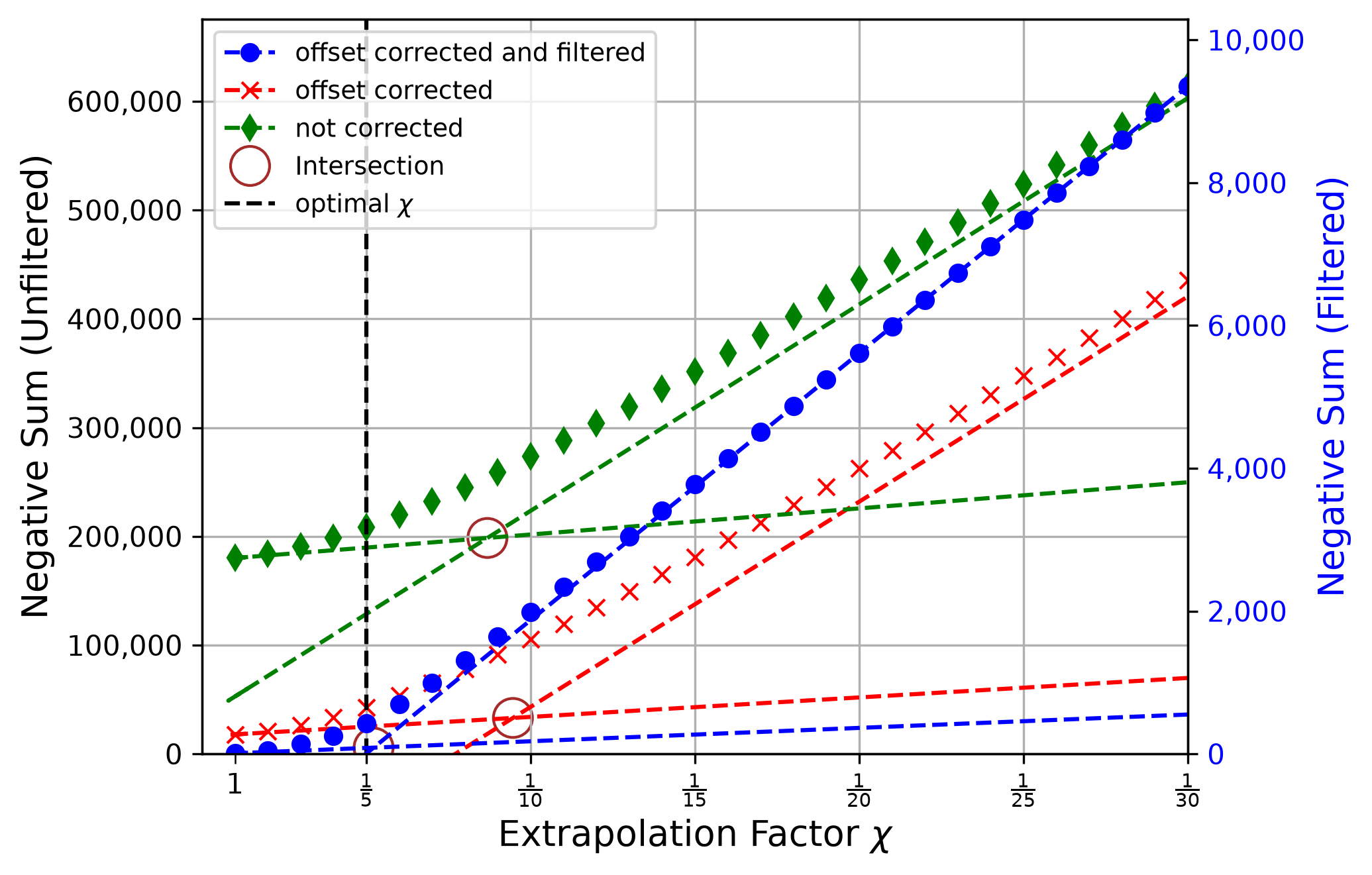}
    \caption{
        \textbf{Showcase example of `Negative Sum Explosion' in action.}
        Negative sum of $\densxtr$ as a function of the inverted extrapolation factor for a simulated dataset (see \autoref{sec:sim-methods}) with ground-truth $\extrapol = 0.2$ under three scenarios.
        Dots are computed values and dashed lines are the under-extrapolation and over-extrapolation fits.
        The first scenario (green) sums over the entire unit cell of $\densxtr$ without the $F_{000}$ correction.
        The second (red) sums over the entire unit cell with the $F_{000}$ correction applied.
        The third (blue) applies the $F_{000}$ correction and restricts the sum to voxels with $\densdiffr < -3\,\rmsddiff$.
        The resulting values are substantially smaller and are plotted on a separate scale (right axis).
        Linear fits are computed over fixed ranges $1 \leq \xtrfacnse \leq 2$ (some points are not plotted, as they increase monotonically) and $100 < \xtrfacnse < 200$, chosen far from any plausible $\extrapol$ to avoid bias in selecting the transition region.
        The intersections yield $\extrapol$ values of $0.12$, $0.11$ and $0.19$ for the three scenarios.
    }
    \label{fig:example_nse}
\end{figure}

\autoref{fig:example_nse} shows how correcting for these two factors affects our simulated dataset (details described in \autoref{sec:sim-methods}).
Restricting the sum to voxels with $\densdiffr < -3\,\rmsddiff$ alongside the $F_{000}$ correction substantially narrows the transition region, and the intersection recovers an estimate close to the ground-truth value.

NSE rests on a clear physical assumption, but its reliability depends on a sharp transition between the under-extrapolation and over-extrapolation regimes.
This transition can be sharpened by carefully selecting the input voxels of $\densxtr$ -- focusing on a region of interest and discarding small values in $\densdiff$ -- and such selection matters more as difference-map quality degrades.

\subsection{Vacuum Level Matching}
\label{sec:vacuum-matching}

These observations motivated us to develop \textit{Vacuum Level Matching}, a modification of the NSE approach.
The goal remains to identify the point at which negative difference density cancels the reference density.
Instead of aggregating over the whole map, we evaluate this cancellation for each voxel individually.
This allows us to assess the reliability of each per-voxel estimate and combine only those that satisfy the selection criteria described below.

This method relies on at least a small fraction of the density in the reference state being replaced by vacuum.
This is a sound assumption for changes inside a protein, where solvent cannot enter or is held in a fixed position.
It does not apply when a ligand binds without displacing any atoms of the reference structure.
At the protein surface and beyond, vacated regions are typically refilled by bulk solvent, so the density drops to the solvent level rather than to vacuum.
Nevertheless, as long as some changes also occur inside the protein, the method can still identify the optimal extrapolation factor, as we argue in \autoref{sec:vm-noise}.

\subsubsection{Local estimates}
\label{sec:vm-noise}

To discover where over-extrapolation has occurred, one wants to focus on voxels corresponding to positions where atoms were present in the reference state but are no longer present in the triggered state.
So, our first goal is to produce per-voxel estimates of extrapolation factors at which the reference density is completely eliminated.

To that end, we start with the general expression for the extrapolated map in \autoref{eq:xtr-general}.
Now we express it instead as an addition of real-space densities $\rho$ rather than complex structure factors $\complex{F}$, yielding
\begin{equation}
    \densxtrr
    = \xtrfacnse\occu\weight
    \left(\denstargetr - \densdarkr + \noisediff(\rr) \right)
    + \densdarkr
    ,
    \label{eq:vm-realspace}
\end{equation}
where we model the effect of processing in two parts: a global scalar rescaling factor $\weight$ (primarily from difference-map processing, including the reference-phase correction prefactor) and per-voxel deviations due to noise or error $\noisediff(\rr)$.

What we can see in the expression above is that we can eliminate the contribution of the reference density when $\xtrfacnse\occu\weight=1$.
For each voxel individually, this expression becomes a linear expression and can easily be solved.
As in the case of NSE, we want to focus on the case where as much reference density as possible is eliminated.
So, we require $\densxtrr=0$, and now obtain an estimate of the extrapolation factor for each voxel.
\begin{equation}
    \xtrestimate
    = \occu\,\weight
    \left(1 - \frac{\denstargetr}{\densdarkr}
    - \frac{\noisediff(\rr)}{\densdarkr}\right)
    = \frac{-\densdiffr}{\densdarkr}.
    \label{eq:vm-noise}
\end{equation}

\begin{figure}[htb]
    \centering
    \includegraphics[width=1\linewidth]{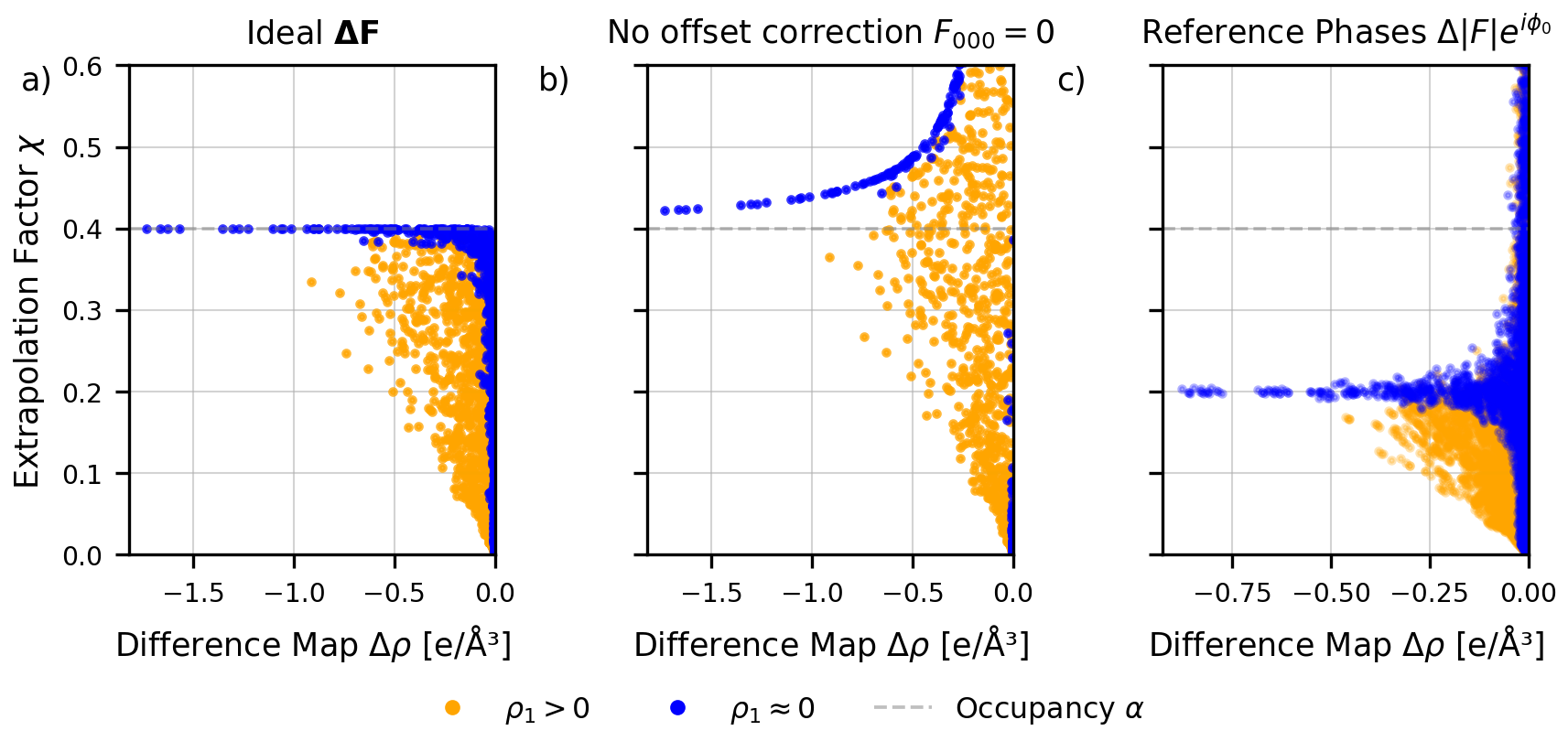}
    \caption{
        \textbf{Idealized illustration of $\mathbf{\xtrestimate}$ under different conditions.}
        Per-voxel estimate $\xtrestimate = -\densdiffr/\densdarkr$ plotted against $\densdiffr$ for a simulated dataset based on rsEGFP2 (PDB 8A6G, modeled with $\occusim = 0.4$).
        (a) Corrected $F_{000}$, exact difference-map phases. The estimates concentrate at and below the true value, with the orange voxels forming the underestimating tail (effect of $\denstargetr$).
        (b) As~(a) but without the $F_{000}$ correction, displacing the estimates above the true value.
        (c) Corrected $F_{000}$, difference map computed from reference-state phases only. The reference-phase prefactor halves the true $\extrapol$ and the difference-map noise disperses the estimates across a broad band.
    }
    \label{fig:cistrans-comparison}
\end{figure}

Most single estimates are biased.
The estimate at a single voxel is reliable only where the reference density dominates both confounding terms, that is, where $\denstargetr \ll \densdarkr$ and $|\noisediff(\rr)| \ll \densdarkr$.
Where this fails, the estimate is systematically biased, towards over-extrapolation for $\denstargetr$ (\autoref{fig:cistrans-comparison}a) and in either direction for noise.

Refilling by bulk solvent is a special case of this bias.
A vacated voxel has $\denstargetr = 0$, whereas a voxel refilled by solvent has $\denstargetr = \denssolvent$, which lowers its estimate by the factor $(1 - \denssolvent/\densdarkr)$.
Crucially, the refilled voxel also has a less negative difference density, since $\densdarkr - \denssolvent < \densdarkr$.
Because the voxels are ranked by $\densdiffr$, most negative first (\autoref{sec:vm-selection}), vacated voxels enter the estimate first and dominate the accepted window.
Refilled voxels that do enter early can only do so at large $\densdarkr$, where $\denssolvent/\densdarkr$, and hence their bias, is small.
Any partial or imperfect refilling lies between these two limits, so the estimate remains approximately correct even when the solvent model is not exact.

\subsubsection{Towards a single estimate and its reliability}
\label{sec:vm-selection}

Equation (\ref{eq:vm-noise}) enables us to compute the extrapolation factor for which the extrapolated map coincides with the vacuum level -- a quantity that varies from voxel to voxel. Our next task is to identify a subset of voxels that provide a consistent estimate of a global extrapolation factor to use. That subset should include a region of the difference map that contains genuine negative signal that exceeds any background noise.

In \autoref{fig:cistrans-comparison}, we plot $\xtrestimate$ as a function of $\densdiff$ for a number of theoretical scenarios, to show how the per-voxel estimate of the extrapolation factor varies as a function of how strongly negative that voxel is in the difference map.
Overall, we see that the different estimates $\xtrestimate$ form a narrow distribution around a single value for sufficiently negative difference-map voxels.
Using reference phases halves the optimal extrapolation factor and introduces noise.
A constant error in reference amplitudes increases in relative strength with decreasing $-\densdiff$.
Also, we showed in \autoref{eq:vm-noise} that a large reference density leads to lower confounding factors.

Therefore, choosing which voxels to include in a set that defines a global extrapolation factor involves an intrinsic bias-variance tradeoff.
Including more voxels suppresses the influence of outliers, but \autoref{eq:vm-noise} shows that including voxels with substantial
$\denstargetr$ introduces a systematic bias towards over-extrapolation.
\begin{figure}[htb]
    \centering
    \includegraphics[width=0.9\linewidth]{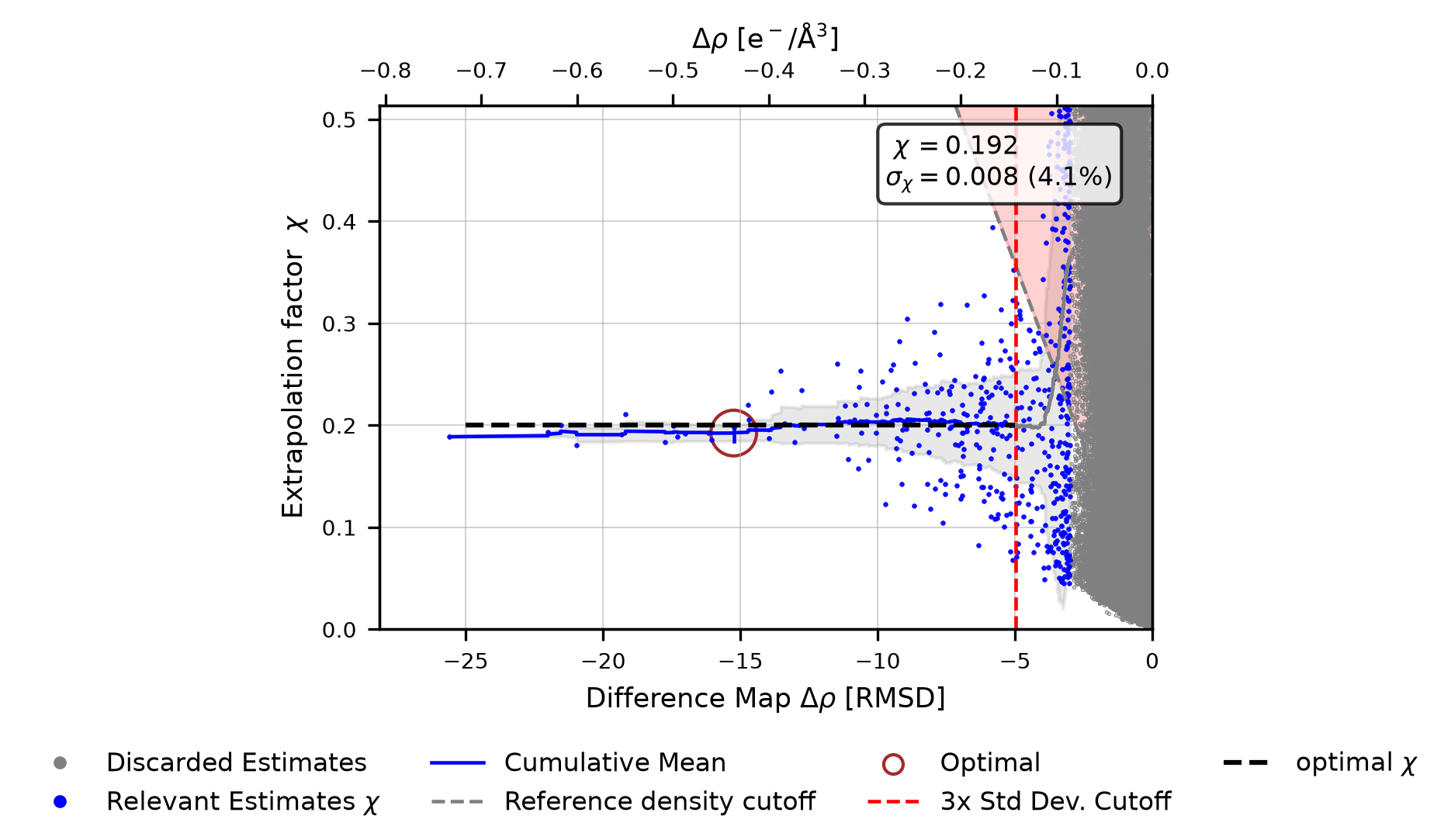}
    \caption{
        \textbf{Showcase example of `Vacuum Level Matching' in action.} Same dataset and noise level as used for the showcases of the preceding methods (rsEGFP2, PDB 8A6G, $\occusim = 0.4$).
        Each blue dot is a per-voxel estimate $\xtrestimate = -\densdiffr/\densdarkr$, plotted against $\densdiffr$ in units of $\rmsddiff$ (lower axis) and in absolute units (upper axis).
        Estimates discarded by the difference-map filter are shown as grey dots.
        The reference density cutoff (grey dashed line) is the line traced by $-\densdiffr/\densdarkcut$, where $\densdarkcut$ is set to the solvent level.
        Any estimate above that line (red background) comes from a small denominator $\densdarkr$ and is correspondingly less reliable.
        The blue line is the running cumulative mean and the grey band its running standard deviation, both accumulated from the most negative $\densdiffr$ towards zero.
        The red dashed line marks the dynamic cutoff, beyond which voxels are excluded from the estimate, and the cumulative mean is shown in grey beyond it.
        The reported estimate (circled, with a blue vertical line) is the median of the running cumulative mean over the accepted window.
    }
    \label{fig:example_vm}
\end{figure}

To balance these considerations, we employ the following heuristic procedure.
We sort the voxels by $\densdiffr$, starting with the most negative, and compute a cumulative mean and standard deviation of $\xtrestimate$ along this ordering.
We set the reference-density cutoff $\densdarkcut$ to the bulk-solvent density, which defines the diagonal boundary $-\densdiffr/\densdarkcut$ in the per-voxel estimate plot.
Accumulation stops when the running mean plus three running standard deviations first exceeds this boundary,
\begin{equation}
    -\densdiffr/\densdarkcut < \langle\xtrestimate\rangle + 3\stdxtr .
\end{equation}
The final estimate of the extrapolation factor is given by the median of the running mean over the accepted set of voxels. As stated, this procedure is a heuristic, and we have not shown it is optimal under any probabilistic model. As we will show, however, in empirical testing it has proven robust and effective.

\autoref{fig:example_vm} shows the method applied to our rsEGFP2 example case, where the running mean returns 0.192 compared to the optimal $\trueextrapol = 0.2$, a 4~\% difference.
Implementation details are given in \autoref{sec:vm-methods}, and in the following section we will further quantify each method's accuracy on simulated data.

NSE shares with this approach more than the intuition of eliminating the reference density.
In fact, as we show in \autoref{sec:nse-vacuum-comp}, the intersection method of NSE, with some mild assumptions, is a $(-\densdiffr)$-weighted mean of the per-voxel quantities
$\xtrfacnse(\rr) = \frac{\densdarkr}{-\densdiffr}$.
NSE therefore rests on the same per-voxel estimates as Vacuum Level Matching, but aggregates them in two ways that ours does not.
It averages in the inverse quantity $\xtrfacnse$ rather than in $\extrapol$, and it takes a $(-\densdiffr)$-weighted mean over whichever voxels are supplied rather than selecting them by the reliability criterion established above.

\section{Comparing \texorpdfstring{$\chi$}{chi}-estimation methods on simulated data}
\label{sec:comparison}

\begin{figure}[htbp]
    \centering
    \includegraphics[width=\linewidth]{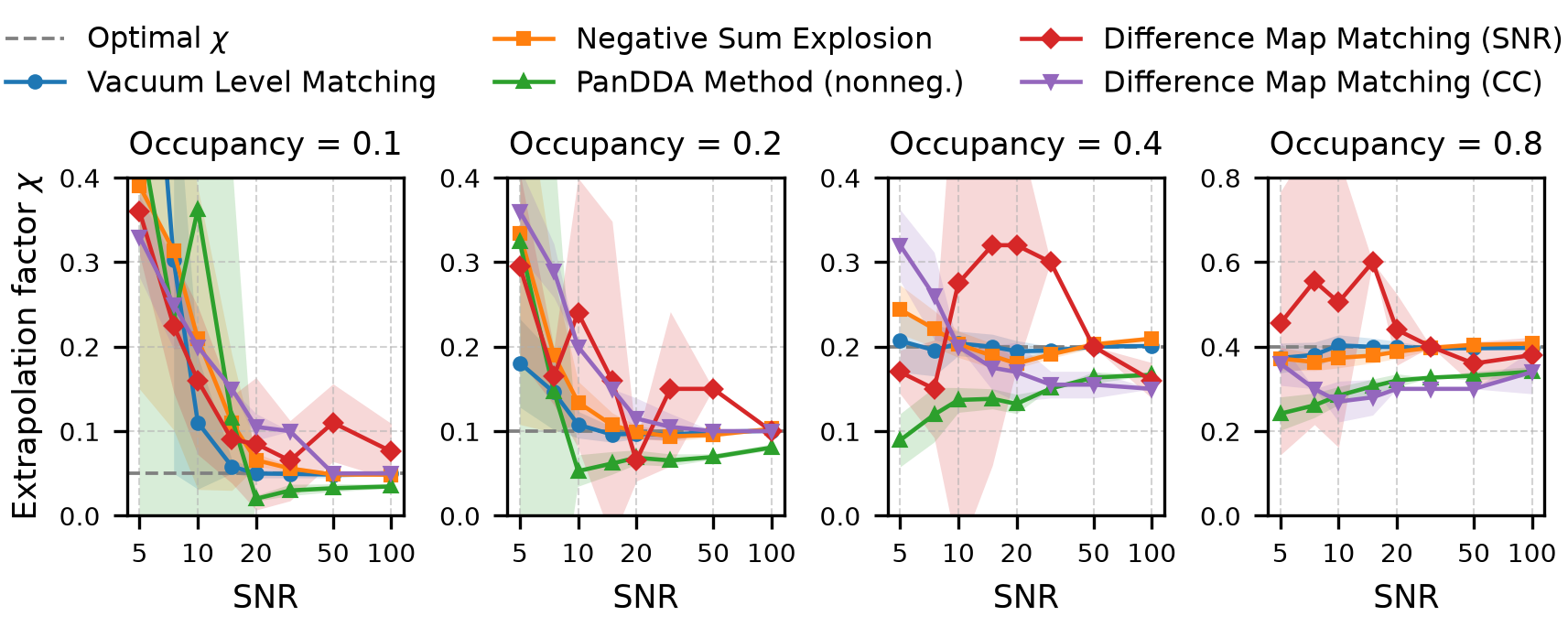}
    \caption{
        \textbf{Comparison of extrapolation factor estimates benchmarked using simulated TR-SFX signals.}
        Estimated extrapolation factor as a function of signal-to-noise ratio $\snr$ for four true occupancies.
        Lines and markers show the mean across ten independent noise samples, and shaded bands show $\pm$ one standard deviation.
        The dashed grey line marks the optimal $\trueextrapol=\occu/2$, the optimal extrapolation factor under the reference-phase approximation (see \autoref{sec:extrapolation-factor}).
    }
    \label{fig:sim-aggregate}
\end{figure}

To systematically compare the methods we introduced in \autoref{sec:xtr-techniques}, we use simulated reference and mixed-state datasets of rsEGFP2 at a range of occupancies and noise levels.
Details of the simulation setup are given in \autoref{sec:sim-methods}.
\autoref{fig:sim-aggregate} shows the predicted extrapolation factor as a function of the signal-to-noise ratio $\snr$ for each of the four occupancies tested.
The estimates from each method are averaged across ten independent noise samples for each data point, and the mean and standard deviation (shaded area) are shown.

The difference maps use reference-state phases, and the structure factors are not processed in any other way.
The optimal extrapolation factor is therefore half the true occupancy, $\occu/2$, a value all methods approximately recover at low noise ($\snr \geq 50$; see \autoref{sec:extrapolation-factor}).
At very low SNR, none is reliable.
Similarly, when occupancy is low, the smaller signal means that, for a given $\snr$, the difference map is noisier, and again all methods struggle.
The methods are therefore best distinguished at moderate noise, where they diverge and the choice between them matters.

The Peak Integration variant of Difference Map Matching has the highest variance, because, as explained in \autoref{sec:x8}, it often reports a very high $\extrapol$ when no good solution is found.
The CC method recovers the correct extrapolation factor at high SNR and low occupancy, but over-extrapolates consistently at higher occupancy.
For both Difference Map Matching routines this may stem from their dependence on the reference-model mismatch $\deltadark$, which varies with occupancy.

The nonnegative PanDDA variant shows a systematic over-extrapolation bias across TRX-like data (the standard version, not shown, even more so).
Once its variance has dropped, its estimate stays below $\trueextrapol$ and approaches it only gradually as the SNR increases.
This systematic offset suggests that the heuristic could be recalibrated to perform more accurately.

Negative Sum Explosion rests on the same physical principle as Vacuum Level Matching, the minimization of negative density in the extrapolated map.
At high SNR it reliably recovers the true extrapolation factor.
In the noisy regime, however, it shows higher variance, since it sums over the whole map and cannot down-weight the voxels where the per-voxel estimate is least reliable.

Vacuum Level Matching converges to the same high-SNR limit, recovering the true value where NSE does.
Its advantage lies in the noisy regime.
By resolving the estimate voxel by voxel and filtering adaptively, it keeps its mean closer to the truth with a smaller spread, and this advantage widens as noise increases.

In this simulation test, Vacuum Level Matching recovers the correct extrapolation factor consistently and accurately up to higher noise levels than the other methods tested.
Simulations are, however, an imperfect proxy for experiment, so we next turn to two real-data test cases.

\section{Vacuum Level Matching on experimental data}
\label{sec:real-data}
Evaluating extrapolation-factor methods on real data is fundamentally different from evaluating them on simulations.
No ground truth is available, and any reliable measure of accuracy would itself constitute a method.
The assumptions inherent to any given method ought to match the experimental reality.
In the case of NSE and Vacuum Level Matching, this means that the largest changes result in a structure where some electron density moves and leaves behind a region free of any electron density.

First, we turn to the rsEGFP2 dataset of \citet{fadiniSerialFemtosecondCrystallography2023}, which was already the basis for our simulation study in the previous section.
As Vacuum Level Matching performed best, it is the only method that we consider in this part.
Following that case study, in \autoref{sec:photolyase-real}, the multi-time-point photolyase data of \citet{christouTimeresolvedCrystallographyCaptures2023} let us test whether estimates remain coherent across preprocessing pipelines and whether the built-in quality metrics correctly separate data that support a confident prediction from those that do not.

\begin{figure}[htb]
    \centering
    \includegraphics[width=0.6\linewidth]{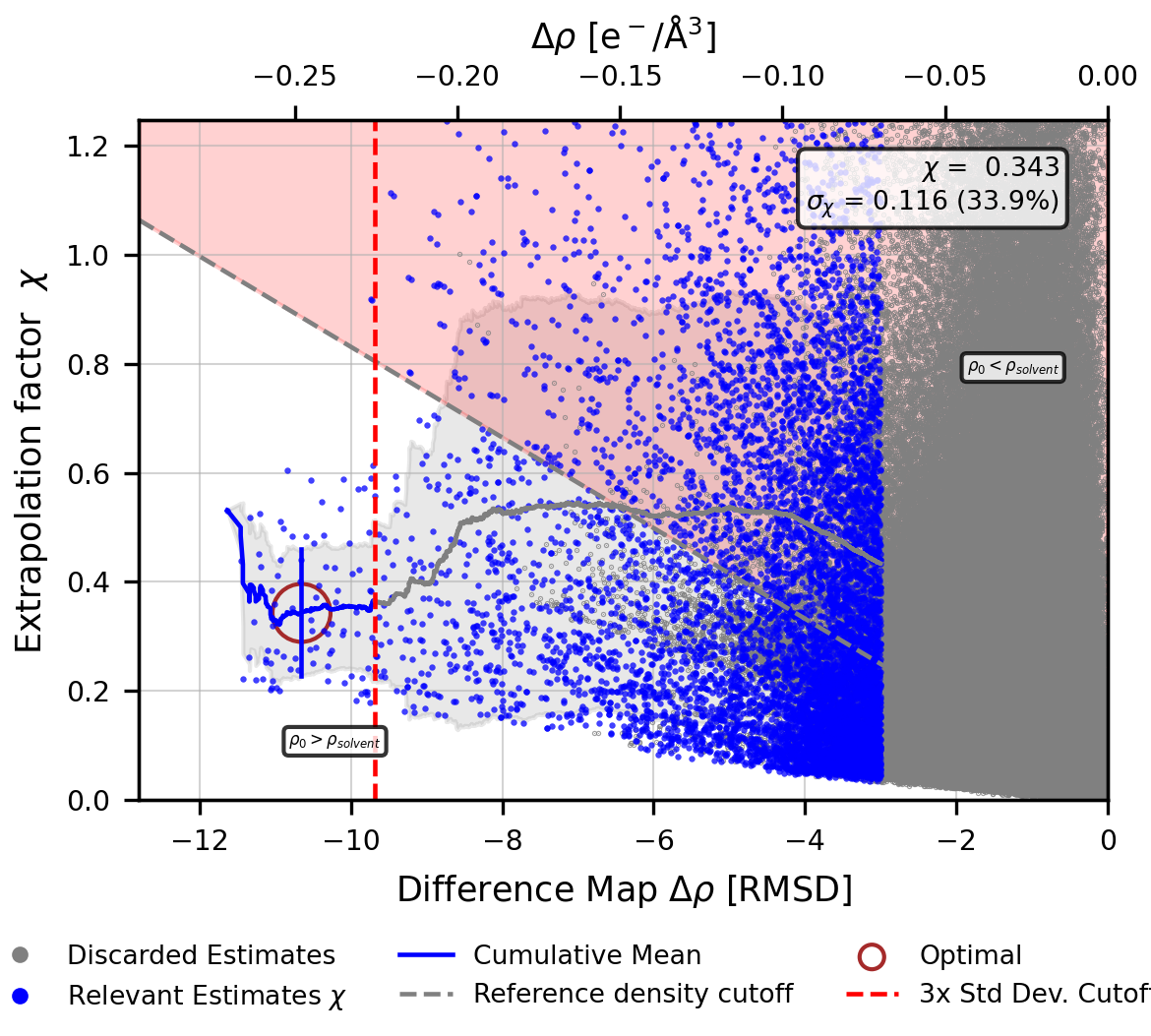}
    \caption{
        \textbf{Vacuum Level Matching on experimental data}.
        The difference map is calculated from the rsEGFP2 dataset of \citet{fadiniSerialFemtosecondCrystallography2023} after applying iterative TV denoising.
        Per-voxel estimates $\xtrestimate = -\densdiffr/\densdarkr$ are accumulated from the most negative $\densdiffr$ towards zero (lower axis in units of $\rmsddiff$, upper axis in absolute units).
        The cumulative mean is drawn in blue, with the shaded band indicating the $\pm 1\stdxtr$ region.
        Beyond the red dashed line, where the cutoff condition $\langle\xtrestimate\rangle + 3\stdxtr > -\densdiffr/\densdarkcut$ is first met, the cumulative mean is shown in grey.
        The red shaded region marks $\densdarkr < \densdarkcut$, where the small denominator amplifies noise in the per-voxel estimate.
    }
    \label{fig:rsEGFP2-plot}
\end{figure}

\subsection{Single-time-point application: rsEGFP2}
\label{sec:rsEGFP2-real}

rsEGFP2 is a reversibly switchable fluorescent protein whose chromophore undergoes a \textit{cis}--\textit{trans} isomerization on illumination.
The structural transition is localized to the chromophore and a few neighboring residues.
The dataset of \citet{fadiniSerialFemtosecondCrystallography2023} captures this transition at low triggered-state occupancy and was used to construct the simulated benchmark of \autoref{sec:comparison}.

We first scale the reference and mixed-state structure factors to the model and calculate the respective mean densities (or zero-frequency terms) \citep{langProteinStructuralEnsembles2014}.
Using the reference model phases, we are able to calculate the reference density $\densdark$ with a simple Fourier transform.
Then, we calculate difference maps between the two, using the reference structure phases and applying iterative TV denoising (IT-TV).
This allows us to obtain the difference density $\densdiff$ with more refined amplitudes and phases.
For more details, we refer to \autoref{sec:exp-methods}.

We show the resulting extrapolation factor estimates $\xtrestimate$ in \autoref{fig:rsEGFP2-plot}.
As expected, there is a narrow-distribution regime for the most negative $\densdiffr$.
The cutoff condition is able to identify this regime and picks the extrapolation factor within the stable regime as predicted.
Compared with the simulated showcase of \autoref{fig:example_vm}, the narrow-distribution regime has a higher variance.
While the difference is visually apparent, to quantify such differences, we consider the relative spread $\stdxtr/\extrapol$ as the most informative measure.
At over 30~\%, it is comparatively large, which reflects the high level of noise inherent to this dataset.
For the same reason, the cutoff condition is already activated at more negative $\densdiffr$.
Therefore, the accepted window spans a shorter range than in the simulated example.
The reported $\extrapol = 0.34$, as we argued above, does not necessarily correspond to the actual occupancy, but ought to lead to the best estimate of the extrapolated density.

\begin{figure}[htbp]
    \centering
    \includegraphics[width=0.6\linewidth]{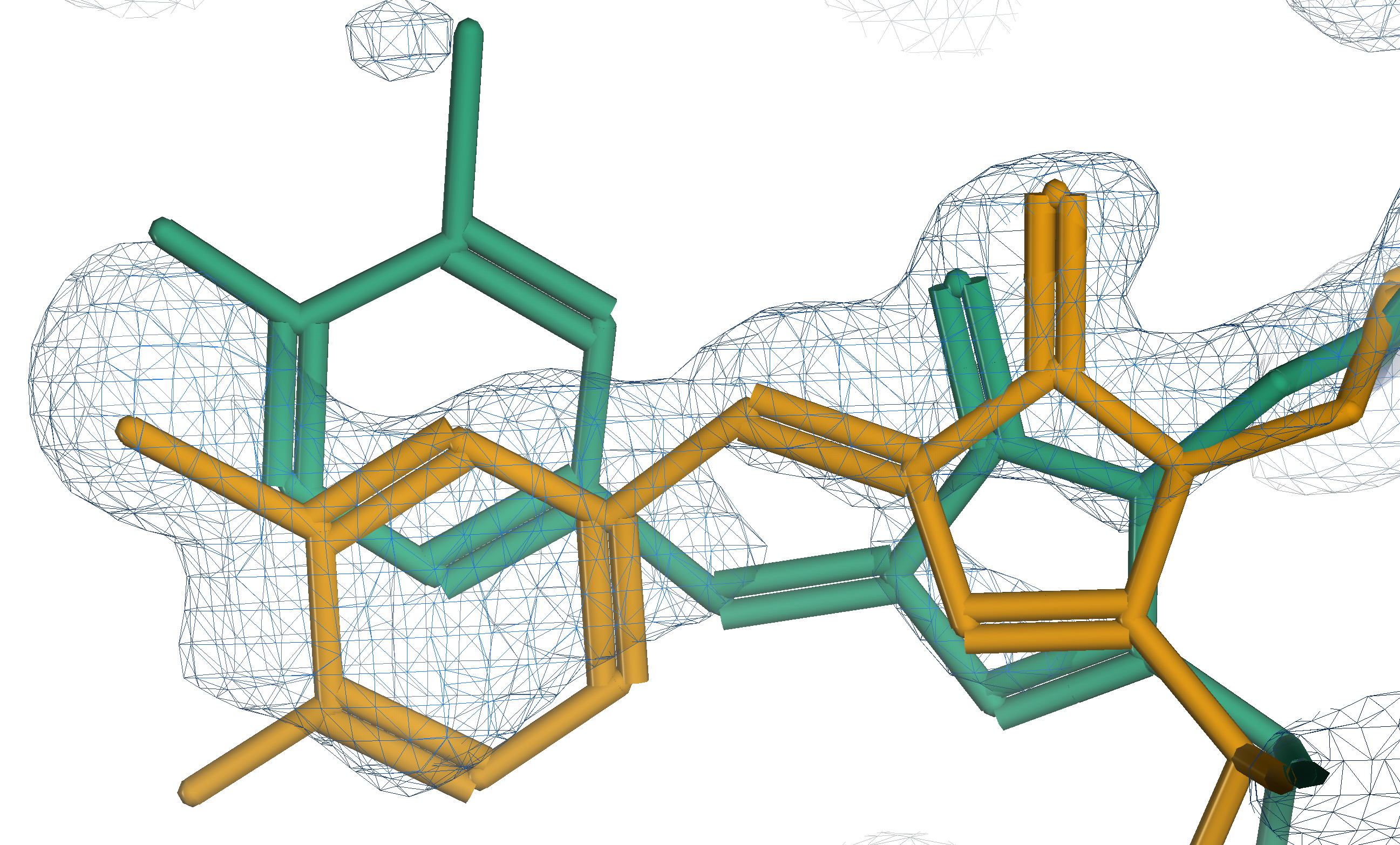}
    \caption{
        \textbf{Example of extrapolated electron density}.
        The rsEGFP2 chromophore is shown with the extrapolated density calculated with the optimal $\extrapol$ given by \autoref{fig:rsEGFP2-plot}.
        The \textit{trans} (reference-state) chromophore model is shown in green, and the \textit{cis} (triggered-state) chromophore model from the published structure is in orange.
        The extrapolated map (mesh) is contoured around the chromophore region at \rmsd{} $= 1$.
    }
    \label{fig:rsEGFP2-coot}
\end{figure}

For a real measurement, the practical test of an extrapolation-factor estimate is whether the resulting map resolves the structural transition the experiment was designed to capture.
\autoref{fig:rsEGFP2-coot} shows the extrapolated electron density at the chromophore site, computed with the $\extrapol$ reported in \autoref{fig:rsEGFP2-plot}.
The map follows the \textit{cis}-chromophore model rather than the \textit{trans} reference state, confirming that the reported $\extrapol$ removes the reference contribution without notably over-extrapolating.

\subsection{Photolyase: multi-time-point consistency across preprocessing}
\label{sec:photolyase-real}
\begin{figure}[htb]
    \centering
    \includegraphics[width=\linewidth]{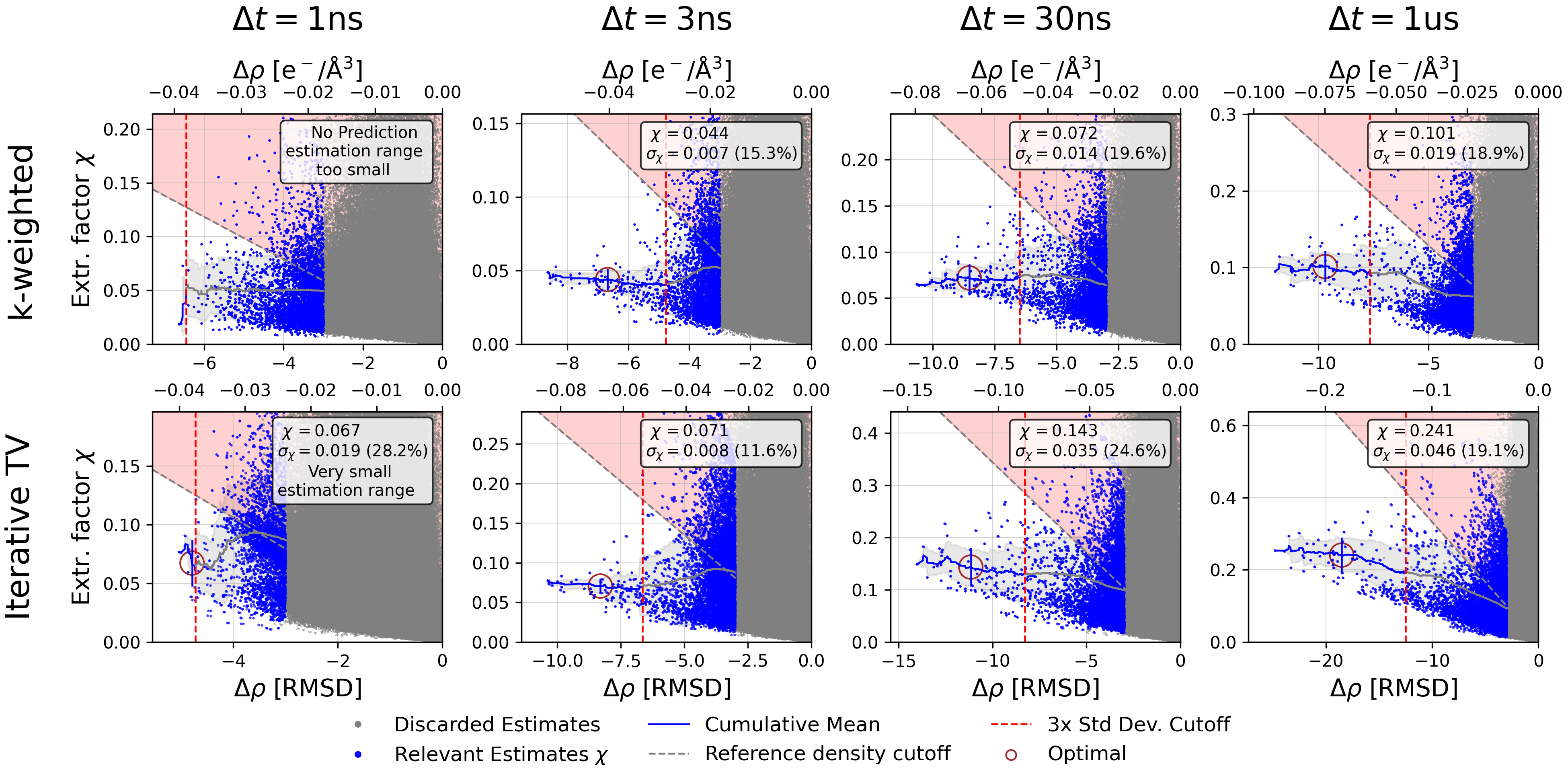}
    \caption{
        \textbf{Vacuum Level Matching applied to a range of difference maps}. Using photolyase data of \citet{christouTimeresolvedCrystallographyCaptures2023}, we show four time points with two difference-map processing pipelines.
        The top row shows \kweighted{} difference maps as published by \citet{christouTimeresolvedCrystallographyCaptures2023}, and the bottom row shows iterative TV-denoised difference maps as described in \autoref{sec:exp-methods}.
        Each panel displays per-voxel estimates (blue dots), the running cumulative mean (solid blue line), the running standard deviation (grey band), the dynamic cutoff (red dashed line), and either a numerical $\extrapol$ prediction or a warning flag.
    }
    \label{fig:photolyase-grid}
\end{figure}

Photolyase is a flavoenzyme that repairs UV-induced cyclobutane pyrimidine dimer (CPD) lesions.
\citet{christouTimeresolvedCrystallographyCaptures2023} presented ten time points spanning picoseconds to hundreds of microseconds.
Within the first few nanoseconds, the cyclobutane ring is cleaved one bond at a time, passing through a single-bond intermediate at around \qty{1}{\ns} and reaching a fully cleaved state within a few nanoseconds.
On the microsecond timescale, the cleaved thymines flip out of the binding pocket to complete product release.

We use these data as a retrospective test of our method.
We show four time points, \qty{1}{\ns}, \qty{3}{\ns}, \qty{30}{\ns} and \qty{1}{\us}, each processed with two different difference-map pipelines.
The first pipeline uses the \kweighted{} difference maps, computed using structure-factor weighting, as published by \citet{christouTimeresolvedCrystallographyCaptures2023}.
The other pipeline is what we consider the current state of the art, iterative TV denoising (IT-TV), which adjusts the map in real space and iteratively aligns these changes with measured structure factors, leading to improved phase estimates.

\begin{figure}[htb]
    \centering
    \includegraphics[width=\linewidth]{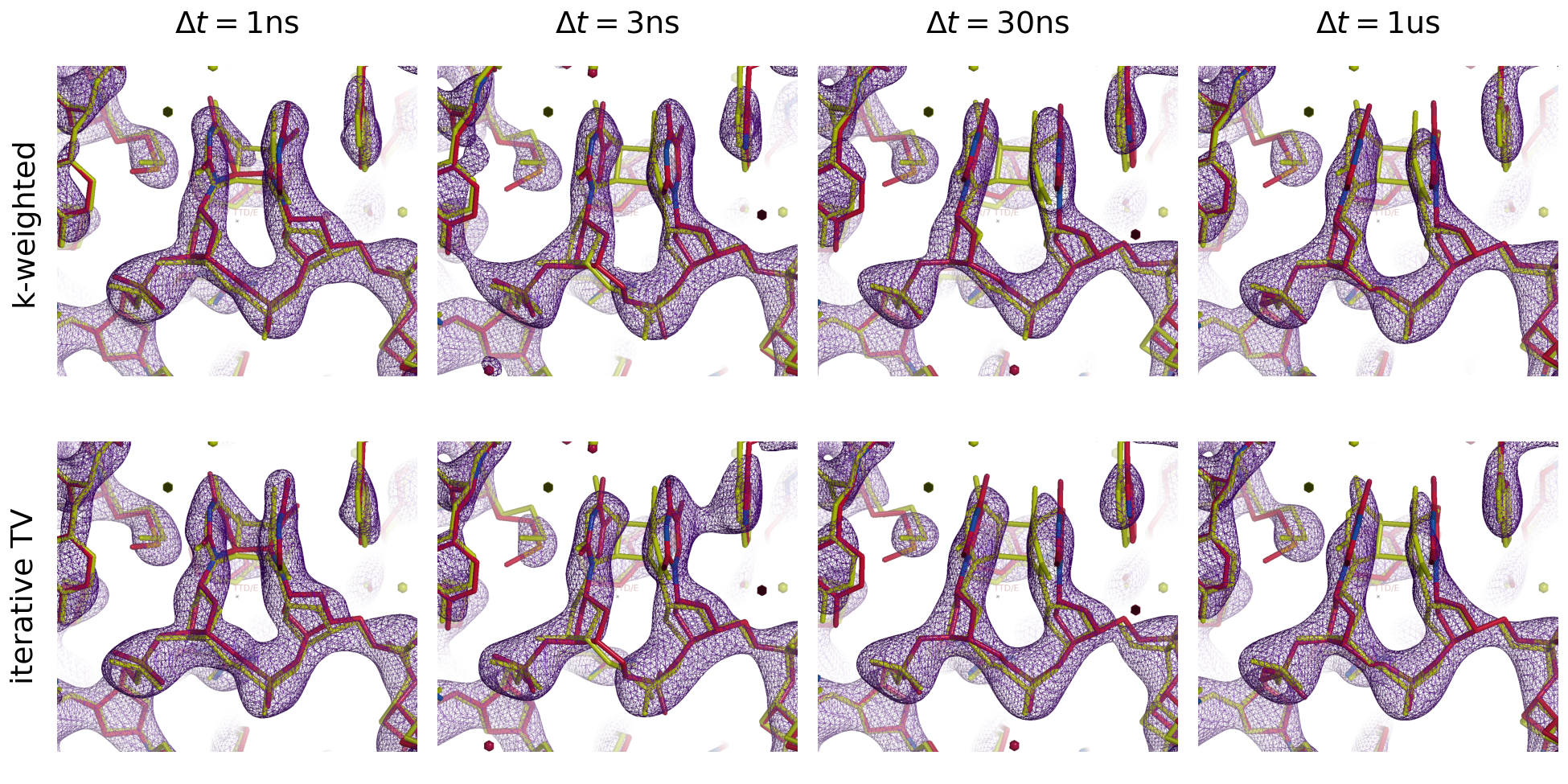}
    \caption{
        \textbf{Extrapolated electron density maps.} Here, we show the main change with extrapolated densities using $\extrapol$ as estimated in \autoref{fig:photolyase-grid}.
        Top row: maps from the \kweighted{} difference map of \citet{christouTimeresolvedCrystallographyCaptures2023}.
        Bottom row: maps from the iterative TV-denoised difference map.
        Both maps are contoured at \rmsd{} $= 2$.
        The reference model (yellow) and the published extrapolated model (red) are displayed as well.
    }
    \label{fig:photolyase-maps}
\end{figure}

\autoref{fig:photolyase-grid} shows the per-voxel distributions and cumulative statistics at these four time points under the two pipelines, and \autoref{fig:photolyase-maps} shows the corresponding extrapolated maps.
We discuss the two together, since the estimate returned for a given time point is only meaningful in light of the map it produces.

At \qty{3}{\ns}, \qty{30}{\ns} and \qty{1}{\us}, the difference-map signal shows a similar rearrangement.
Consequently, under both pipelines Vacuum Level Matching finds a stable running mean (\autoref{fig:photolyase-grid}).
For these three time points, we observe a wide range of difference-map voxels that share a similar $\extrapol$.
Compared with the rsEGFP2 dataset of \autoref{sec:rsEGFP2-real}, where the signal-dominated region was shorter, the photolyase maps stay narrow over a much wider span of $\densdiff$.
This increases the confidence that we have identified a credible extrapolation factor.

The absolute $\extrapol$ values differ between \kweighted{} and IT-TV at every time point, and the two pipelines disagree by different amounts at different time points.
Yet the corresponding maps in \autoref{fig:photolyase-maps} all resolve the same cleavage-region structure.
Differences between time points may be physical, reflecting pump-laser power or the actual occupancy of the state.
But the discrepancy between \kweighted{} and IT-TV at a single time point arises from the same underlying data and the same structural state, so it can only be processing-induced.
Processing alone can therefore scramble the relationship between physical occupancy and extrapolation factor.
This is consistent with the analysis of \autoref{sec:extrapolation-factor} and reinforces the recommendation that any reported $\extrapol$ be accompanied by the pipeline that produced it.

Our automated safeguards raise a warning at the \qty{1}{\ns} time point.
They detect that only a small number of estimates were included before the cutoff rule was activated.
Thus, there are too few data points to proceed in an automated manner.

Here, this happens because the difference map is weaker overall, and its most negative voxels do not sit at the two main cleavage sites as they do at later time points.
Instead, they form a collection of small shifts distributed around that region.
The per-voxel estimates are correspondingly widely spread, so under default settings the warning conditions trigger for both pipelines, though with different severity, because the pipelines suppress noise differently.
For IT-TV, a warning is raised because the accepted window is short, spanning less than 20~\% of the negative-$\densdiff$ range.
For the \kweighted{} map, very few data points fall within a sufficiently small standard deviation, and no estimate is reported at all.
In both cases, the method signals that it cannot itself guarantee a sound estimate and that visual inspection is required.

In this case, inspection of the plot still yields a usable value.
The running cumulative mean is approximately flat over a substantial portion of the most negative voxels, so a defensible estimate can be read off with the caveat noted.
We chose $\extrapol = 0.05$ for the \kweighted{} map and $0.067$ for IT-TV, around which the mean hovers down to roughly $4.2~\rmsddiff$.
Both produce sensible extrapolated maps (\autoref{fig:photolyase-maps}), with the IT-TV maps resolving the cleavage-region density in somewhat greater detail, consistent with the real-space sharpening that the IT-TV procedure applies.

We obtained $\extrapol$ estimates for all four time points under two distinct preprocessing pipelines, and the resulting extrapolated maps resolve the structural changes that previously required manual intervention.
Taken together, the photolyase results show that Vacuum Level Matching produces internally consistent estimates within a pipeline and across structurally similar time points, that the absolute values depend on the preprocessing pipeline in the manner anticipated by \autoref{sec:extrapolation-factor}, and that, depending on data quality, the method either reports automatically or correctly defers to human review.

\section{Conclusion}
TRX is rapidly maturing, yet the accurate separation of mixed structural states remains a major bottleneck.
Tools such as Xtrapol8 now handle all steps from measured structure factors to a first refined model, putting TRX analysis within reach of non-specialists.
This makes it all the more important that each step, including extrapolation-factor estimation, is reliable and transparent about its assumptions.
We argued, and demonstrated on an experimental example, that extrapolation-factor estimates cannot simply be equated with occupancies, because difference-map processing affects them in ways that are not generally invertible.
In the photolyase data, the same physical state under two processing pipelines yields different extrapolation factors, both of which produce interpretable results.
We surveyed the methods in current use, and the negative-density assumption of \citet{tripathiPHDependencePhotoactive2012} inspired us to introduce Vacuum Level Matching.
This is a method for extrapolation-factor estimation that is both more transparent and more accurate than existing approaches, as we show in \autoref{sec:comparison}.

A central design goal was transparency.
The method not only produces an estimate but also communicates how dependable that estimate is, surfacing the per-voxel evidence and flagging cases where the data do not support a confident prediction.
Its output falls into three diagnostic states.
When no warning fires, the estimate is reported as confident.
When a warning fires, the estimate is flagged for manual evaluation. As the photolyase results show, a defensible value can often still be read off the per-voxel plot in this case.
When no estimate satisfies the selection criteria, the method returns an error rather than a number.
These states are built around several failure modes identified and characterized in this work (\autoref{sec:vm-selection}).
Untested data regimes may reveal further modes.
The conservative diagnostic criteria are intended to flag unanticipated failures or prevent an estimate from being returned, reducing the risk of reporting an unreliable estimate as confident.

Several limitations bind the method's scope.
Vacuum Level Matching rests on the two-state approximation.
It estimates a single extrapolation factor for the dominant negative-density signal in a map; it does not determine how many distinct structural changes contribute to that signal.
The estimate therefore describes the dominant state and may not apply to all changed regions.
When several changes can be isolated spatially, local masks allow each region to be analyzed separately.
When the contributing states overlap spatially, dedicated multi-state separation is required beyond what a single-map method can provide.
The method also requires substantial negative difference density associated with the displacement of existing density.
Its applicability to ligand binding is therefore limited to cases where the ligand displaces existing density rather than only occupying empty space.

Both experimental datasets, and the simulation derived from one of them, involve small, spatially localized changes: a chromophore isomerization in rsEGFP2 and bond cleavage in photolyase.
These test cases do not probe a large conformational rearrangement, a membrane protein, or a system in which the difference signal is delocalized across the unit cell.
A next step would be to investigate how this method handles large-scale changes and to deploy it on a broader set of time-resolved data.

Another next step is to refine the heuristics that locate the narrow-distribution regime of $\xtrestimate$ and determine $\extrapol$ from it (\autoref{sec:vm-selection}).
Although effective in our tests, these heuristics lack a formal statistical basis.
One route is to estimate the noise level of the difference map, which would give the expected distribution of per-voxel estimates for genuinely vacated voxels.
This would allow a more principled classification of plausible vacuum-state values and indicate how much of the variance in the narrow-distribution regime is explained by noise alone.
Similarly, the matched baseline could be made local.
At present, the method only requires that at least some voxels are vacated to vacuum, and voxels refilled by solvent are effectively outweighed (\autoref{sec:vm-noise}).
Matching each voxel against its expected local background instead -- vacuum inside the protein, the solvent level at its surface -- would let solvent-refilled voxels contribute to the estimate rather than merely being tolerated, increasing the number of informative voxels.

Beyond the method itself, we would advise against using the difference map to determine occupancy directly, particularly when the effect of processing on the occupancy--extrapolation-factor relationship cannot be quantified.
We instead recommend first using the extrapolation factor to obtain a best estimate of $\complex{\strucxtr}$.
The structure factors calculated from a model refined against that estimate can then be substituted for $\complex{\structarget}$ in \autoref{eq:fmix-complex}.
With a single unknown left, one can estimate the occupancy $\occu$ that best explains the measured amplitudes,
$\strucmix \approx \bigl|(1-\occu)\,\complex{\strucdark} + \occu\,\complex{\strucxtr}\bigr|$,
with $\complex{\strucdark}$ and $\complex{\strucxtr}$ held fixed.

\section{Methods}
\subsection{Simulation of rsEGFP2}
\label{sec:sim-methods}

\paragraph{Generation of structure factors.}
To benchmark the methods of \autoref{sec:xtr-techniques} against a known ground truth, we generated simulated reference and mixed-state datasets of rsEGFP2 with a chromophore undergoing \textit{trans}-to-\textit{cis} isomerization on illumination.
As a basis, we used PDB entry 8A6G, which contains both the \textit{trans} and \textit{cis} states at partial occupancies.
The reference state is modeled as the \textit{trans} state at full occupancy and the mixed state as the \textit{cis} state at the specified occupancy $\occusim$ and the \textit{trans} state at $1-\occusim$, with all atoms outside the chromophore held at their reference positions.
Complex structure factors for both states were computed with the \textit{gemmi} package \citep{wojdyrGEMMILibraryStructural2022}, including a bulk-solvent contribution, at a resolution of 1.6~\AA.

\paragraph{Simulating noise.}
We model two sources of error in the structure-factor amplitudes, following the combination of a counting term and a term proportional to the intensity commonly used in error models for scaling, motivated by the error model used in ref. \citep{evansScalingAssessmentData2006} (see Appendix A3).
First, serial data are dominated by multiplicative errors, such as partiality, shot-to-shot variation in pulse intensity and crystal size, and imperfect scaling, which perturb each intensity in proportion to its magnitude \citep{kirianFemtosecondProteinNanocrystallography2010, whiteCrystallographicDataProcessing2013}.
Propagated to amplitudes, these give a noise term proportional to $|F|$.
Second, photon counts follow a Poisson distribution, so $\sigma_I = \sqrt{I}$.
Propagating this uncertainty via $\sigma_{|F|} \approx \sigma_I/(2|F|)$ gives an approximately constant noise level on the amplitudes that is most consequential for the weakest reflections.
We represent this constant term by the average amplitude $\langle F_\text{clean}\rangle_\resring$ over the outermost, highest-resolution reflection ring $\resring$, defined as all $\qq$ within $0.1$~\AA\ of the resolution cutoff of 1.6~\AA.
To the structure factors of each model, we apply noise as follows:
\begin{equation}
    F_\text{noisy}(\qq) = F_\text{clean}(\qq) + \frac{1}{\snr} \left(\gausnoise_1 F_\text{clean}(\qq) + \gausnoise_2\langle F_\text{clean}\rangle_\resring \right),
\end{equation}
where $\gausnoise_1$ and $\gausnoise_2$ are random numbers drawn from a Gaussian distribution with zero mean and unit standard deviation.
We refer to this scaling parameter as $\snr$.
Because the additive term is fixed at the average amplitude of the highest-resolution ring, $\snr$ is an upper bound on the signal-to-noise ratio for all but the weakest reflections.

\paragraph{Configuration of extrapolation factor estimation methods.}
We apply Difference Map Matching via Xtrapol8 using the \texttt{fextr} and \texttt{fofo} options, which apply no structure-factor reweighting, and throughout evaluate Difference Map Matching at 15 values of $\extrapol$ between 0.01 and 0.9.
We report both the Peak Integration version -- combining positive and negative peaks -- and the CC version.
For the Peak Integration method, we report the maximum for which $\densdiffxtr$ and $\densdifffofo$ agree in sign at the largest absolute voxels of $\densdifffofo$ (\autoref{sec:x8}), which excludes the spurious high scores of under-extrapolated maps at large $\extrapol$.
The CC method needs no such check, since a sign inversion directly lowers the Pearson correlation.

We use the NSE variant with the difference map filtered at $3\rmsddiff$ and the $F_{000}$ offset included, as discussed in \autoref{sec:nse}.
To fit the under-extrapolation regime, we calculate $\densxtr$ with $\xtrfacnse$ at five equidistant points between 1 and 2; to fit the over-extrapolation regime, we use five equidistant points between 100 and 200.
The fits shown in \autoref{fig:example_nse} are also based on these calculations.

For PanDDA, we use the nonnegative formulation from \autoref{sec:pandda}.
Vacuum Level Matching is performed as specified in \autoref{sec:vacuum-matching}.

\paragraph{Simulation design.}
For the benchmark of \autoref{sec:comparison}, we evaluated every method on a common grid of occupancies ($\occusim \in \{0.1, 0.2, 0.4, 0.8\}$) and noise levels ($\snr \in \{5,\allowbreak 7.5,\allowbreak 10,\allowbreak 15,\allowbreak 20,\allowbreak 30,\allowbreak 50,\allowbreak 100\}$).
Each (occupancy, noise) combination was repeated over ten independent noise samples, and all methods were applied to the identical difference map for each sample, so that the spread in the results reflects the methods rather than the data.
The showcase dataset, displayed throughout \autoref{sec:xtr-techniques}, uses $\occusim = 0.4$, giving an optimal $\extrapol = 0.2$, and $\snr=15$.

\subsection{Experimental data processing}
\label{sec:exp-methods}

\paragraph{Determining the constant component.}
Because the zero-frequency term $F_{000}$ cannot in general be measured, the reconstructed maps lack a constant offset (see \autoref{sec:nse}).
We restore this offset following \citet{langProteinStructuralEnsembles2014}.
The absolute density of the reference map is the sum of a protein contribution and a bulk-solvent contribution.
First, the structure factors of a model map are calculated using \textit{gemmi}.
Then, protein and solvent regions are identified with a mask.
We estimate the bulk-solvent density $\denssolvent$ by a bounded one-dimensional search.
For each candidate $\denssolvent$, we add a flat density over the solvent mask to the model map.
We scale the resulting model to the experimental data and compute the mean absolute amplitude difference, restricted to reflections coarser than $4$~\AA, where the bulk solvent dominates.
The chosen $\denssolvent$ is the one that minimizes this difference.
The protein contribution is the volume-weighted mean density of the model map over the protein region, and the solvent contribution is the bulk-solvent level $\denssolvent$ over the solvent mask.

\paragraph{Difference Maps.}
In \autoref{sec:real-data}, we reuse the \kweighted{} difference maps reported by \citet{christouTimeresolvedCrystallographyCaptures2023}, calculated using Xtrapol8 with $k=1$.
The second method that we use is iterative TV denoising \citep{fadiniDenoisingIterativePhase2025} (meteor version 0.5.0, available at \url{https://github.com/rs-station/meteor}).
\paragraph{Full extrapolation pipeline.}
Starting from the experimentally determined structure factors $\strucdark$ and $\strucmix$ together with a reference model, we first scale $\strucdark$ and $\strucmix$ to the model.
A difference map is then computed following either the \kweighted{} or the iterative TV denoising approach, and the constant component $F_{000}$ is determined for the reference structure.
We next estimate the optimal extrapolation factor $\extrapol$ (\autoref{sec:xtr-techniques}).
Finally, we extrapolate following \autoref{eq:xtr-complex}, restated here for convenience,
\begin{equation*}
    \complex{\strucxtr} = \complex{\strucdark} + \xtrfacnse\,\complexstrucdiff.
\end{equation*}

\subsection{Vacuum Level Matching implementation details}
\label{sec:vm-methods}

\paragraph{Sampling rate.}
Real-space difference maps were sampled on a grid with a sampling factor of five, that is, $2.5\times$ the Nyquist--Shannon minimum, to provide adequate spatial resolution for the per-voxel analysis.
Finer sampling does not add information.
It samples the same map more densely and thereby increases the number of points entering the running statistics.
Finer sampling is therefore useful when the cumulative mean is unstable, and the sampling can be reduced when the statistics are unaffected.

\paragraph{Filtering the difference map.}
We restrict the analysis to voxels that are themselves strongly negative, $-\densdiffr \geq 3\,\rmsddiff$, and connected into a volume of at least $V_\text{min} = 1.5~\text{\AA}^3$.
This removes isolated negative outliers that are likely to be noise rather than genuine structural signal.
The threshold $V_\text{min}$ is deliberately permissive.
When the map contains extensive negative signal, a more restrictive $V_\text{min}$ may be preferable.

\paragraph{Reference-density cutoff.}
The main text (\autoref{sec:vm-selection}) halts the accumulation using the diagonal $-\densdiffr/\densdarkcut$.
We set $\densdarkcut = \denssolvent$, the bulk-solvent level, determined during preprocessing while estimating $F_{000}$.
This choice is motivated on two grounds.
First, as noted in \autoref{sec:vm-selection}, a small $\densdarkr$ amplifies noise in the numerator of $\xtrestimate$, and $\denssolvent$ marks the density below which this amplification dominates.
Second, a voxel with $\densdarkr$ at or below $\denssolvent$ is indistinguishable from bulk solvent, so a negative $\densdiffr$ there is as plausibly attributable to a solvent fluctuation as to a genuine change in ordered protein density.
Such voxels carry no reliable structural information regardless of their denominator.

\paragraph{Warning criteria.}
Three checks lower the confidence in an automatic estimate.
In each case, the data violate an assumption underlying the method.
The first raises a warning when the accepted window spans less than 20~\% of the negative-$\densdiff$ range, indicating that the narrow-distribution regime covers too small a range for a reliable estimate.
The second raises a warning when the running mean is unstable across the window, defined as the minimum running mean (excluding the first few voxels) falling below two thirds of its maximum.
The third raises a warning when the running standard deviation is large relative to the reported estimate.
When any condition is met, the estimate should not be trusted without inspection.

\clearpage

\appendix
\newcommand{\strucdelta}{F_{1-0}}
\newcommand{\phasedelta}{\phase_{1-0}}
\section{Connection between Vacuum Level Matching and NSE}
\label{sec:nse-vacuum-comp}

This appendix shows that, under two mild assumptions, the Negative Sum Explosion is equivalent to a weighted average of the per-voxel zero-crossings that define Vacuum Level Matching.
The NSE estimate is the intersection of two regression lines, one fitted in the under-extrapolation regime (small $\xtrfacnse$) and one in the over-extrapolation regime (large $\xtrfacnse$).
At each $\xtrfacnse$, the negative sum collects every voxel for which $\densxtrr<0$.
To write each regression line as a linear equation, we first establish which voxels contribute in each regime.

We consider only voxels that can physically cross zero.
Voxels with $\densdarkr \leq 0$ are excluded, since reference density below vacuum is unphysical; such voxels arise mainly from the missing constant offset and are almost entirely removed once $F_{000}$ is restored (\autoref{sec:nse}).
Voxels with $\densdiffr \geq 0$ are excluded as well: given $\densdarkr > 0$, their extrapolated density $\densdarkr + \xtrfacnse\densdiffr$ could only become negative for $\xtrfacnse < 0$, that is, by subtracting the difference rather than adding it.
As in the main text, we further restrict the analysis to sufficiently negative difference density, giving the voxel set
\begin{equation}
    \rrneg = \{\rr : \densdiffr \leq -c,\; \densdarkr > 0\}, \qquad c \geq 0.
\end{equation}
Setting $c = 0$ includes all negative-difference voxels, whereas $c > 0$ (in the main text, $c = 3\,\rmsddiff$) excludes voxels whose small $|\densdiffr|$ would delay their zero-crossing to very large $\xtrfacnse$.

The first assumption concerns the under-extrapolation regime.
For sufficiently small $\xtrfacnse$, every voxel satisfies $\densdarkr > -\xtrfacnse\densdiffr$, so $\densxtrr$ stays positive and the negative sum is flat at zero.
Its regression line is therefore zero, and the intersection of the two lines lies at the root of the over-extrapolation line.

The second assumption concerns the over-extrapolation regime.
For sufficiently large $\xtrfacnse$, the difference term dominates the reference in every voxel of $\rrneg$, so all of them have crossed zero.
The negative sum then runs over the fixed set $\rrneg$, independent of $\xtrfacnse$, and is linear in $\xtrfacnse$,
\begin{equation}
    \sum_{\rr\in\rrneg}\densxtrr
    = \sum_{\rr\in\rrneg}\densdarkr
    + \xtrfacnse \sum_{\rr\in\rrneg}\densdiffr ,
\end{equation}
so the regression recovers this line exactly.
We find its root by dividing the intercept by the slope,
\begin{equation}
    \xtrfacnse
    = \frac{\sum_{\rr\in\rrneg} \densdarkr}
           {{\sum_{\rr\in\rrneg} -\densdiffr}}.
\end{equation}
Having found the global estimate, we turn our attention to the per-voxel estimates.
The same construction, applied to a single voxel, gives an estimate at each point,
\begin{equation}
    \Tilde\xtrfacnse(\rr) = \frac{\densdarkr}{-\densdiffr}.
\end{equation}
Taking the average over all per-voxel crossing points, weighted by $-\densdiffr$, we obtain
\begin{align}
    \begin{split}
        \xtrfacnse
        &= \frac{\sum_{\rr\in\rrneg} \Tilde\xtrfacnse(\rr) (-\densdiffr)}
                {{\sum_{\rr\in\rrneg} -\densdiffr}} \\
        &= \frac{\sum_{\rr\in\rrneg} \frac{\densdarkr}{-\densdiffr}(-\densdiffr)}
                {{\sum_{\rr\in\rrneg} -\densdiffr}} \\
        &= \frac{\sum_{\rr\in\rrneg} \densdarkr}
                {{\sum_{\rr\in\rrneg} -\densdiffr}}.
    \end{split}
\end{align}
This is the same expression as the global estimate of $\xtrfacnse$.
In this scenario, NSE is therefore a weighted average of the per-voxel zero-crossings that Vacuum Level Matching makes explicit, with each voxel weighted by $-\densdiffr$.
If either assumption is violated, for example if the low-$\xtrfacnse$ regime is not flat at zero or some voxels in $\rrneg$ have not yet crossed zero within the fitting range, the equivalence is only approximate and the two estimates differ slightly.

Because NSE accesses this average through a linear regression rather than directly, the choice of fitting range is arbitrary, whereas Vacuum Level Matching exposes the per-voxel distribution and the influence of each choice. The weighting itself, as a function of difference density, is not very different once the analysis is restricted to strongly negative voxels.
The difference-map noise enters \autoref{eq:vm-noise} as an additive term, $\occu\,\weight\,\noisediff/\densdarkr$, whose magnitude is set by $\densdarkr$ and is independent of the true $\extrapol$ at the voxel.
Under the reciprocal map $\xtrfacnse = 1/\extrapol$, an additive perturbation $\delta$ on $\extrapol$ becomes a perturbation of magnitude $\delta/\extrapol^{2}$ on $\xtrfacnse$, whose size depends on the value of $\extrapol$ at the voxel.
Aggregating in $\extrapol$ therefore preserves a uniform additive noise scale across voxels, whereas aggregating in $\xtrfacnse$, as NSE does implicitly, does not.

\section{Occupancy variance in serial crystallography}
\label{sec:occupancy-variance}

In a serial femtosecond crystallography (SFX) experiment, the assumption of a uniform, constant occupancy is almost always violated. In reality, the occupancy $\occu$ changes from crystal to crystal due to a variety of experimental factors, among them the size of the individual crystal, its position within the liquid jet, its alignment relative to the focused optical laser beam, and the orientation of the crystal lattice relative to the laser polarization.

Because SFX data processing merges diffraction patterns from thousands of individual crystals, the analysis ultimately obtains an average intensity. To understand the implications, we examine how the variance in occupancy across the ensemble affects the final structure-factor amplitudes.

To make the dependence on $\occu$ directly visible, we introduce $\complex{\strucdelta} = \complex{\structarget} - \complex{\strucdark}$, the pure complex structural difference between the triggered state (state 1) and the reference state (state 0).
In the main text, we primarily talk about the observed difference electron density $\complex{\strucdiff} = \occu\,\complex{\strucdelta}$.
Yet, given that our goal is to understand the role of $\occu$, we make it explicit in the following derivations.
If we consider a single crystal (or a single ``shot'') with a specific occupancy $\occu$, the measured intensity is
\begin{align}
    I_{mix}(\occu) &= |\complex{\strucdark} + \occu \complex{\strucdelta}|^2 \nonumber \\
    &= |\complex{\strucdark}|^2 + 2\occu \Re(\complex{\strucdark}^* \complex{\strucdelta}) + \occu^2 |\complex{\strucdelta}|^2 .
\end{align}

Because the dataset is an aggregate, we calculate the expected value (the average measurement) across the ensemble. Using the statistical identity $\langle\occu^2\rangle = \langle\occu\rangle^2 + \sigma_\occu^2$, where $\langle\occu\rangle$ is the mean occupancy and $\sigma_\occu^2$ is its variance, the average intensity becomes
\begin{align}
    \label{eq:intensity-variance}
    \begin{split}
        \langle I_{mix} \rangle &= |\complex{\strucdark}|^2 + 2\langle\occu\rangle \Re(\complex{\strucdark}^* \complex{\strucdelta}) + (\langle\occu\rangle^2 + \sigma_\occu^2) |\complex{\strucdelta}|^2 \\
        &=
        |\complex{\strucdark} + \langle\occu\rangle \complex{\strucdelta}|^2 + \sigma_\occu^2 |\complex{\strucdelta}|^2
        .
    \end{split}
\end{align}
Here $\sigma_\occu$ is the spread in occupancy across crystals and is unrelated to the difference-map and running-estimate standard deviations used in the main text.

To obtain the observed mixed-state amplitude $|F_\mathrm{mix}|$, we take the square root of the average intensity.
Writing $\phase_0$ and $\phasedelta$ for the phases of $\complex{\strucdark}$ and $\complex{\strucdelta}$, respectively, gives
\begin{align}
    |F_{mix}| &= \sqrt{\langle I_{mix} \rangle} \nonumber \\
    &= |\complex{\strucdark}| \Biggl[
    1 + 2\langle\occu\rangle
    \frac{|\complex{\strucdelta}|}{|\complex{\strucdark}|}
    \cos(\phase_0 - \phasedelta) \nonumber \\
    &\qquad + (\langle\occu\rangle^2 + \sigma_\occu^2)
    \frac{|\complex{\strucdelta}|^2}{|\complex{\strucdark}|^2}
    \Biggr]^{1/2}.
\end{align}
For reflections where the occupancy-weighted structural differences are small relative to the reference amplitude, we expand consistently through second order in the structural difference.
Using $\sqrt{1+x} = 1 + x/2 - x^2/8 + \cdots$, we obtain
\begin{align}
    |F_{mix}| &\approx |\complex{\strucdark}|
    + \langle\occu\rangle |\complex{\strucdelta}|
    \cos(\phase_0 - \phasedelta) \nonumber \\
    &\quad + \frac{(\langle\occu\rangle^2 + \sigma_\occu^2)
        |\complex{\strucdelta}|^2}{2|\complex{\strucdark}|} \nonumber \\
    &\quad - \frac{\langle\occu\rangle^2 |\complex{\strucdelta}|^2
        \cos^2(\phase_0 - \phasedelta)}{2|\complex{\strucdark}|}.
\end{align}
Subtracting the reference amplitude and combining the quadratic terms yields
\begin{align}
    \label{eq:final-diff}
    \Delta |F| &= |F_{mix}| - |\complex{\strucdark}| \nonumber \\
    &\approx \langle\occu\rangle |\complex{\strucdelta}|
    \cos(\phase_0 - \phasedelta) \nonumber \\
    &\quad + \frac{|\complex{\strucdelta}|^2}{2|\complex{\strucdark}|}
    \left[\sigma_\occu^2 + \langle\occu\rangle^2
    \sin^2(\phase_0 - \phasedelta)\right].
\end{align}
Thus, occupancy variance contributes an additional positive term to the difference amplitude at second order.
Even without occupancy variance, a second-order correction remains when the structural difference has a component perpendicular to the reference structure factor in the complex plane.

Setting $\sigma_\occu = 0$ and retaining only the first-order term recovers the single-occupancy approximation used in \autoref{sec:xtr-darkphases}:
\begin{equation*}
    \Delta |F| \approx \langle\occu\rangle |\complex{\strucdelta}|
    \cos(\phase_0 - \phasedelta).
\end{equation*}
Assigning the reference phase and using
\begin{equation*}
    \cos(\phase_0 - \phasedelta)\,e^{i\phase_0}
    = \tfrac12 e^{i\phasedelta}
    + \tfrac12 e^{i(2\phase_0 - \phasedelta)}
\end{equation*}
reproduces the first-order decomposition in \autoref{eq:fxtr-complex}, with $\occu = \langle\occu\rangle$.

\section{Decomposition of the Difference Map Matching score}
\label{sec:diffmap-matching-derivation}

We derive the decomposition in \autoref{eqn:diffmap-matching} and its consequence for the CC method.
We consider positive extrapolation factors and represent post-processing by $\weightobs$ and $\weightxtr$.

Substituting the extrapolated map into the processed difference map and adding and subtracting the observed difference term gives
\begin{align}
    \densdiffxtr
    &= \weightxtr\left[
    \densdark^\text{obs} + \xtrfacnse(\densmix-\densdark)
    - \densdark^\text{calc}\right] \nonumber \\
    &= \xtrfacnse\weightobs(\densmix-\densdark)
    + \xtrfacnse(\weightxtr-\weightobs)(\densmix-\densdark) \nonumber \\
    &\quad + \weightxtr(\densdark^\text{obs}-\densdark^\text{calc}) \nonumber \\
    &= \xtrfacnse\densdifffofo + \xtrfacnse\deltadiff + \deltadark,
    \label{eq:diffmap-matching-expansion}
\end{align}
where $\deltadiff=(\weightxtr-\weightobs)(\densmix-\densdark)$ is the mismatch between the two processing steps and $\deltadark=\weightxtr(\densdark^\text{obs}-\densdark^\text{calc})$ is the reference-model mismatch.

Pearson correlation is invariant to positive rescaling.
Thus, for a fixed set of voxels on which both maps have nonzero variance,
\begin{equation}
    \operatorname{CC}(\densdifffofo,\densdiffxtr)
    = \operatorname{CC}\left(\densdifffofo,
    \densdifffofo+\deltadiff+\extrapol\deltadark\right).
    \label{eq:diffmap-matching-cc}
\end{equation}

Note that if $\deltadark=0$, the score is independent of $\extrapol$; if both mismatch terms vanish, it is unity for every candidate.

Consequently, a CC maximum reflects the pattern introduced by the mismatch terms, rather than a minimum of their absolute magnitude alone.

\begin{funding}
    We acknowledge Deutsches Elektronen-Synchrotron (DESY), a member of the Helmholtz Association, and DASHH, Data Science in Hamburg - Helmholtz Graduate School for the Structure of Matter, for financial support. Parts of this research were funded by and carried out at the Hamburg University of Technology. This work was further
    supported by the Cluster of Excellence ‘CUI: Advanced Imaging of Matter’ of the Deutsche Forschungsgemeinschaft (DFG) – EXC 2056 – project ID 390715994.
\end{funding}

\ConflictsOfInterest{
    There are no conflicts of interest.
}

\DataAvailability{
    The code and the data needed to recreate the plots can be found at
    \url{https://github.com/cvazz/xtr-estimator/} and are available under the Apache 2 license.
}

\bibliography{zotero}
\end{document}